\documentclass[pre,floatfix,longbibliography,superscriptaddress,onecolumn,final,notitlepage]{revtex4-2}
\usepackage[utf8]{inputenc}
\usepackage[T1]{fontenc}
\usepackage{multirow}
\usepackage{tabularx}
\usepackage{graphicx,caption}
\usepackage{amsfonts,amsmath,amssymb}
\usepackage{comment}
\usepackage{lineno}
\usepackage{hyperref}
\usepackage[explicit]{titlesec}

\usepackage[nameinlink,capitalize]{cleveref}

\newcommand{\Exp}{\mathbb{E}}

\newcommand{\Var}{\mathrm{Var}}
\newcommand{\Cov}{\mathrm{Cov}}

\newcommand{\be}{\begin{equation}}
\newcommand{\ee}{\end{equation}}
\newcommand{\bea}{\begin{eqnarray}}
\newcommand{\eea}{\end{eqnarray}}

\newcommand{\ra}{\rightarrow}
\newcommand{\f}[2]{\frac{#1}{#2}}

\newcommand{\ccup}[1]{\left\{#1\right\}}

\newcommand{\bup}[1]{\left(#1\right)}
\newcommand{\rup}[1]{\left[#1\right]}

\renewcommand{\ref}[1]{[\ref{#1}]}

\usepackage{tabularx}

\usepackage{amsfonts,amsmath,amssymb}   
\usepackage{comment}
\usepackage{lineno}
\usepackage{hyperref}
\usepackage[explicit]{titlesec}
\usepackage{graphicx}
\usepackage[labelfont=bf]{caption}
\usepackage[format=hang]{subcaption}
\usepackage{algorithm,algorithmic}
\usepackage{amsmath,fixmath}

\usepackage[algoruled,algo2e]{algorithm2e}
\usepackage{sidecap}
\usepackage{boldline}
\usepackage{setspace}
\usepackage{listings}
\usepackage{fancyvrb}
\usepackage{soul}

\usepackage{mathtools}

\usepackage[table]{}
\usepackage{array}

\usepackage[utf8]{inputenc}

\usepackage{booktabs}
\usepackage{adjustbox}

\usepackage{changepage}

\usepackage[usenames,dvipsnames]{xcolor}
\definecolor{shadecolor}{gray}{0.9}
\usepackage{tabu}

\usepackage[normalem]{ulem} 

\newcommand{\mt}{\mbox{{\small MT}}}
\newcommand{\crep}{\mbox{{\small CRep}}}
\newcommand{\jcrep}{\mbox{{\small JointCRep}}}
\newcommand{\rec}{$\mathsf{r}$}

\begin{document}

\title{Community detection and reciprocity in networks by jointly modeling pairs of edges}

\author{Martina Contisciani}
	\email{martina.contisciani@tuebingen.mpg.de}
	\affiliation{Max Planck Institute for Intelligent Systems, Cyber Valley, Tuebingen 72076, Germany}

\author{Hadiseh Safdari}
	\email{hadiseh.safdari@tuebingen.mpg.de}
	\affiliation{Max Planck Institute for Intelligent Systems, Cyber Valley, Tuebingen 72076, Germany}

\author{Caterina De Bacco}
	\email{caterina.debacco@tuebingen.mpg.de}
	\affiliation{Max Planck Institute for Intelligent Systems, Cyber Valley, Tuebingen 72076, Germany}



\begin{abstract} 
To unravel the driving patterns of networks, the most popular models rely on community detection algorithms. However, these approaches are generally unable to reproduce  the structural features of the network. Therefore, attempts  are always made  to develop  models that incorporate  these network properties beside the community structure.  
In this work, we present a probabilistic generative model and an efficient algorithm to both perform community detection and capture reciprocity in networks. Our approach jointly models pairs of edges with exact 2-edge joint distributions. In addition, it provides closed-form analytical expressions for both marginal and conditional distributions. We validate our model on synthetic data in recovering communities, edge prediction tasks, and generating synthetic networks that replicate the reciprocity values observed in real networks. We also highlight these findings on two real datasets that are relevant for social scientists and behavioral ecologists. Our method overcomes the limitations of both standard algorithms and recent models that incorporate reciprocity through a pseudo-likelihood approximation. The inference of the model parameters is implemented by the efficient and scalable expectation-maximization algorithm, as it exploits the sparsity of the dataset.   We provide an open-source implementation of the code online.
\end{abstract}

\maketitle

\section{Introduction} 
Network models are powerful and flexible tools for representing complex interactions between individual elements in many different fields \cite{fell2000small, newman2001structure, watts1998collective, williams2000simple}.
For instance, in social support networks, each individual is a person or the representative of a household, and each link, tie or arc represents the presence or intensity of a relationship between two individuals. Understanding what core patterns drive the observed set of interactions is of high relevance for scientists and practitioners willing to fully exploit the increased availability of networked datasets.
A popular approach to modeling networks is that of generative models, in particular latent variable models \cite{goldenberg2010}. They are probabilistic models that introduce latent variables to incorporate domain knowledge and capture complex interactions. Of particular interest, is the possibility of recovering clusters of individuals that behave similarly, a problem named community detection \cite{fortunato2010community}. In this framework, the latent variables represent the nodes' community memberships and the structure of interactions between communities, and the aim is to infer these quantities from the data \cite{ball2011efficient,debacco2017multitensor}.  Despite their flexibility and computational efficiency, these models have a main flaw: they fail in reproducing important structural network properties  such as transitivity, reciprocity, or triadic closure \cite{seshadhri2020impossibility,safdari2020generative,peixoto2022disentangling}. Synthetic networks generated from these models tend to have significantly lower values of these properties than those observed in real networks. 
 
One possible reason of this problem is the common assumption of conditional independence: conditioned on the latent variables, network edges are independent and the joint probability distribution is factorized accordingly. This means that an interaction from node $i$ to node $j$ is not directly affected by the interaction in the opposite direction, i.e., the edge $j \rightarrow i$. In latent variable models with community structure, such as the stochastic block model \cite{holland1983stochastic} and its variants, these two edges are fully explained by the membership of the two nodes and sometimes by additional parameters such as degree corrections \cite{karrer2011stochastic}. 
While this assumption has been used to obtain tractable problems, it can be too restrictive in certain real scenarios where non-trivial interaction patterns are observed. For instance, in social support networks, it is likely that the existence of interactions from individual $i$ to individual $j$ does not depend only on the groups that $i$ and $j$ belong to, but also on the fact that $j$ has already previously helped $i$. This tendency of forming mutual connections is called reciprocity \cite{wasserman1994social} and it is an important feature in social networks \cite{ready_power_2021,de2021latent}, journal citations \cite{li2019reciprocity} and email communications \cite{garlaschelli2004patterns,newman2002email}, to name a few. While exponential random graph models can represent such network properties in some form \cite{holland1981exponential,park2004statistical,robins2007introduction,snijders2006new}, they do not incorporate a priori latent variables as community membership. In the previous example, incorporating both community structure and the structural property of reciprocity would help us to understand how an individual interacts with others.  Hence, there is a need  to incorporate both these phenomena within a unique probabilistic framework. 

Recently, \citet{safdari2020generative} tackled this problem by modeling the conditional distribution of \textit{pairs} of edges between the same nodes, an assumption also shared by seminal works \cite{holland1983stochastic,wasserman1987stochastic}. \citet{safdari2020generative} include both communities and reciprocity effects inside the likelihood distribution of the network. This resulted in networks samples with values of reciprocity more similar to those of real data, and better edge predictions. However, this model relies on a pseudo-likelihood approximation for parameters' inference, as the model only specifies conditional distributions, but not the \textit{joint} distribution of a pair of edges. As a result of this approximation, the model is not robust in community detection in the regime where reciprocity plays a role. 
\citet{peixoto2022disentangling} has shown similar results in terms of  triadic closure with a model based on Bayesian inference that combines community structure and this network property. This model  also assumes conditional independence among edges and models conditional distributions of triadic edges.

Here we propose a model that takes into account community structure and reciprocity by specifying a closed-form joint distribution of a pair of network edges, which does not involve approximations. To estimate the likelihood of network ties, we use a bivariate Bernoulli distribution--a special case of the multivariate Bernoulli distribution--where the log-odds are linked to community memberships, and pair-interaction variables. Although these patterns are indicative of two distinct mechanisms of network formation, namely, community structure, and reciprocity, it is reasonable to expect that they are related to each other. For instance, i) the preferred connection between nodes of the same community can induce the presence of reciprocated edges involving similar nodes, and ii) the tendency of forming mutual connections can induce the formation of groups of nodes. This conflation means that we cannot reliably interpret the underlying mechanisms of network formation merely from the abundance of reciprocated edges or observed community structure in network data. Our model takes advantage of the useful properties of the bivariate Bernoulli distribution, i.e., the independence and the uncorrelatedness of the component random variables are equivalent and both the marginal and conditional distributions still follow the Bernoulli distribution. Hence, our model has closed-form analytical expressions and enables practitioners to address with more accuracy questions that were not fully captured by standard models; for instance, predicting the joint existence of mutual ties between pairs of nodes. In addition, its algorithmic implementation is efficient and scalable to large system size, as it exploits the sparsity of network datasets, thus allowing its broad applications across disciplines, e.g., citation networks or neuronal networks that consist of several thousand of nodes.

\section{The model}
The main goal of this work is to develop a probabilistic generative model with latent variables that better captures real scenarios where non-trivial interaction patterns are observed in networks. This is achieved by modeling \textit{jointly} the edges between the same pair of nodes, differently from standard models that assume their conditional independence given the latent variables.  Formally, we model the interactions of $N$ individuals as a binary asymmetric matrix $A$, with entries $A_{ij}$ defining the presence or the absence of connections from node $i$ to node $j$. Our model considers jointly the pair $A_{(ij)}:=(A_{ij}, A_{ji})$ distributed with a bivariate Bernoulli distribution of parameters $\boldsymbol \Theta$, which takes values from $(0, 0), (0, 1),(1, 0)$, and $(1, 1)$ in the Cartesian product space $\{0, 1\}^2 = \{0, 1\}\times\{0, 1\}$.  Its probability density function can be written as
\begin{align}\label{eqn:initial_joint}
\! P(A_{(ij)}|\boldsymbol \Theta) &= P(A_{ij},A_{ji}|\boldsymbol \Theta)\\
\! \nonumber &= p_{11}^{A_{ij}A_{ji}}p_{10}^{A_{ij}(1-A_{ji})}p_{01}^{(1-A_{ij})A_{ji}}p_{00}^{(1-A_{ij})(1-A_{ji})}\\
\!\nonumber &=\f{\exp\ccup{A_{ij}f_{ij} + A_{ji}f_{ji} + A_{ij}A_{ji}J_{(ij)}}}{Z_{(ij)}}, \quad
\end{align}
where $Z_{(ij)}$ is a normalization constant and $p_{00}=1/Z_{(ij)}$. In addition, $p_{00}+p_{10}+p_{01}+p_{11}=1$, and
\be\label{eqn:ftot} 
f_{ij} = \log\bup{\f{p_{10}}{p_{00}}}  ,\,
f_{ji} = \log\bup{\f{p_{01}}{p_{00}}} , \,
J_{(ij)} = \log\bup{\f{p_{11}p_{00}}{p_{10}p_{01}}}  \ .
\ee
Thus, $P(A_{ij}, A_{ji}|\boldsymbol \Theta)$ can be viewed as a member of the exponential family, and can be represented in a log-linear formulation as in \Cref{eqn:initial_joint}, where $f_{ij}, f_{ji}$, and $J_{(ij)}$ represent the natural parameters. $J_{(ij)}$ is called cross-product ratio between $A_{ij}$ and $A_{ji}$ and represents the log-odds of the model. Similar to the Ising model \cite{ising1925beitrag}, if $J_{(ij)}=0$ then the components of the bivariate Bernoulli random vector $(A_{ij}, A_{ji})$ are independent, thanks to the equivalence of independence and uncorrelatedness for multivariate Bernoulli distributions \cite{dai2013multivariate}. In this case, the resulting model would be equivalent to consider the product of two independent Bernoulli distributions. Another interesting property of the bivariate Bernoulli is that both marginal and conditional distributions are univariate Bernoulli. Thus, our model has closed-form equations for joint, conditional and marginal distributions.

We now assume that a set of latent variables capture hidden patterns of the data. There are many possibilities to add these variables: one could act directly on the marginal or conditional first moments, as well as modelling separately the different $p_{\alpha \beta}$, with $\alpha,\beta \in \{0,1\}$. However, we model the log-ratios to ease interpretability and the analytical computations. Specifically, we assume
\begin{align}
f_{ij} &= \log \lambda_{ij} \label{eqn:fij} \\
f_{ji} &= \log \lambda_{ji} \label{eqn:fji} \\
J_{(ij)} &= \log \eta \label{eqn:Jij} \ ,
\end{align}
where 
\be\label{eqn:lambda}
\lambda_{ij} = \sum_{k, q=1}^{K} u_{ik}v_{jq} w_{kq}
\ee
captures mixed-membership community structure as in \citet{debacco2017multitensor} and $\eta$ is the pair-interaction coefficient. The parameters $u_{ik}, v_{jq}$ are entries of $K-$dimensional vectors $\boldsymbol{u_i}$ and $\boldsymbol{v_i}$, the out-going and in-coming communities respectively; and $w_{kq}$ are the entries of a $K\times K$ affinity matrix, which regulates the structure of communities, e.g., assortative when its diagonal entries are greater than off-diagonal entries (homophily). Thus, $\boldsymbol \Theta=(u, v, w, \eta)$ are the latent parameters we want to infer. Through \Crefrange{eqn:fij}{eqn:Jij} we encode the assumptions that community structure drives the process of edge formation, and the edges of a pair of nodes depend on each other explicitly according to the parameter $\eta$. When $J_{(ij)}=0$, the probability of $A_{(ij)}$ is given by the agreements of the communities of $i$ and $j$ only; while a positive value for the log-odds will boost the chance to observe a tie between them. Conversely, $J_{(ij)}<0$ decreases the value of $p_{11}$, the probability that both edges exist. Considering \Cref{eqn:Jij}: $0<\eta<1$ and $\eta>1$ codify a negative and positive interaction between $i$ and $j$, respectively. The first lowers the probability of observing both ties $i\ra j$ and $j\ra i$, while the latter increases it. Finally, $\eta=1$ implies no interaction between $A_{ij}$ and $A_{ji}$. With this model at hand we can estimate observable quantities, valuable for practitioners. For instance, one can ask about the expected value of a given tie in general or conditioned on the existence of the opposite one, quantities defined as:
\begin{align}
\Exp\rup{A_{ij}}&= \f{\lambda_{ij}+\eta \lambda_{ij}\lambda_{ji}}{Z_{(ij)}} \ ,\label{eqn:marginalmean}\\
\Exp\rup{A_{ij}|A_{ji}} &=\f{\eta^{A_{ji}}\lambda_{ij}}{\eta^{A_{ji}}\lambda_{ij}+1} \ ,
\end{align}
and similar for $\Exp\rup{A_{ji}}$ and $\Exp\rup{A_{ji}|A_{ij}}$, see \Cref{app:derivations}. With these quantities one can perform edge prediction tasks, which is crucial when we are limited to a subset of the dataset.

\section{Inference}
We infer the parameters using a maximum likelihood approach. Specifically, we maximize the log-likelihood
\begin{align}
\! \mathcal{L} (\boldsymbol \Theta) &= \sum_{i,j} f_{ij} A_{ij} + \f{1}{2}\sum_{i,j}J_{(ij)}A_{ij}A_{ji}-\f{1}{2}\sum_{i,j}\log Z_{(ij)}\label{eqn:loglik} 
\end{align}
with respect to $\boldsymbol \Theta=(u, v, w, \eta)$. Adopting a variational approach, this is equivalent to maximize
\begin{align}
\! \mathcal{L} (\rho,\boldsymbol \Theta) &=\sum_{i,j} \Bigg[ A_{ij} \sum_{k,q}\rho_{ijkq} \log\Big( \f{u_{ik}v_{jq}w_{kq}}{\rho_{ijkq}}\Big)+  \f{1}{2}A_{ij}A_{ji} \log \eta \\
\! \nonumber&- \f{1}{2}\log\Big(\sum_{k,q}u_{ik}v_{jq}w_{kq} + \sum_{k,q}u_{jk}v_{iq}w_{kq} + \eta \sum_{k,q}u_{ik}v_{jq}w_{kq}\sum_{k,q}u_{jk}v_{iq}w_{kq}+1 \Big)\Bigg] \label{eqn:loglikJensen}\ ,
\end{align}
where we introduced the variational distribution $\rho_{ijkq}$ over the parameters and used Jensen's inequality. The equivalence holds when 
\be
\rho_{ijkq}=\f{u_{ik}v_{jq}w_{kq}}{\sum_{k,q}u_{ik}v_{jq}w_{kq}} \label{eqn:rho}\ .
\ee
We estimate the parameters by using an expectation-maximation (EM) algorithm where at each step one updates $\rho$ using \Cref{eqn:rho} (E-step) and then maximizes $\mathcal{L}(\rho,\boldsymbol \Theta)$ with respect to $\boldsymbol \Theta=(u, v, w, \eta)$ by setting partial derivatives to zero (M-step). This iteration is repeated until the log-likelihood converges. The exact equations for the updates of the parameters are in \Cref{app:derivations}, and the whole routine is described in \Cref{alg:EM}. This algorithm is computationally efficient and scalable to large system sizes as it exploits the sparsity of the dataset. Indeed, all the updates involved in the numerator sum over $A_{ij}$, hence only the non-zero entries count, giving an algorithmic complexity of $O(M\,K^2)$, where $M = \sum_{i,j}A_{ij}$ is the number of ties.

\setlength{\textfloatsep}{5pt}
\begin{algorithm}[H]
\SetKwInOut{Input}{Input}
	\setstretch{0.7}
	\Input{network $A=\{A_{ij}\}_{i,j=1}^{N}$, number of communities $K$.}
  	\BlankLine
	\KwOut{membership matrices $u=\rup{u_{ik}},\, v=\rup{v_{ik}}$; network-affinity matrix $w=\rup{w_{kq}}$; pair interaction parameter $\eta$.}
	\BlankLine
	 Initialize $u,v,w,\eta$ at random. 
	 \BlankLine
	 Repeat until $\mathcal{L}$ convergences:
	 \BlankLine
	\quad 1. Calculate $\rho$ (E-step): 
	\be
	  \rho_{ijkq}=\f{u_{ik}v_{jq}w_{kq}}{\sum_{k,q}u_{ik}v_{jq}w_{kq}} \nonumber
	\ee
	\BlankLine
	\quad 2. Update parameters $\boldsymbol \Theta$ (M-step):  
	\BlankLine
	\quad \quad \quad 
		i) for each pair ($i, k$) update memberships:
		\begin{align}
		\quad u_{ik} &=\f{ \sum_{j,q}A_{ij}\rho_{ijkq}}{\sum_{j}\rup{\f{\sum_{q}v_{jq}w_{kq}(1+\eta \lambda_{ji})}{\lambda_{ij}+\lambda_{ji}+\eta \lambda_{ij}\lambda_{ji}+1}}}  \nonumber \\
		\quad v_{ik} &=\f{ \sum_{j,q}A_{ji}\rho_{jiqk}}{\sum_{j}\rup{\f{\sum_{q}u_{jq}w_{qk}(1+\eta \lambda_{ij})}{\lambda_{ij}+\lambda_{ji}+\eta \lambda_{ij}\lambda_{ji}+1}}} \nonumber
		\end{align}
	\BlankLine
	\quad \quad \quad
		ii) for each pair $(k,q)$ update affinity matrix:
		\be
		\quad w_{kq} =\f{ \sum_{i,j}A_{ij}\rho_{ijkq}}{\sum_{i,j}\rup{\f{u_{ik}v_{jq}(1+\eta \lambda_{ji})}{\lambda_{ij}+\lambda_{ji}+\eta \lambda_{ij}\lambda_{ji}+1}}} \nonumber
		 \ee
	\BlankLine
	\quad \quad \quad
		iii) update pair-interaction parameter:
		\be
		\quad \eta = \f{\sum_{i,j}A_{ij}A_{ji}}{\sum_{i,j}\rup{\f{\lambda_{ij}\lambda_{ji}}{\lambda_{ij}+\lambda_{ji}+\eta  \lambda_{ij}\lambda_{ji}+1}}} \nonumber
		 \ee
	\caption{\jcrep:  EM algorithm}
	\label{alg:EM}
\end{algorithm}

Our model (\jcrep) aims to generalize the method presented in \citet{safdari2020generative} (\crep), which was of inspiration for the latent variables underlying the generative process. We refer to \cite{safdari2020generative} for a detailed explanation of this method and summarize its main properties in \Cref{tab:crepvsjcrep}.

\begin{table*}[ht!]
\caption{\label{tab:crepvsjcrep} \textbf{Properties of \jcrep, \crep, and \mt\text{} models}. $\lambda$ represents the community effect and $\eta$ is the parameter linked to the reciprocity \rec. \mt\text{} is a community detection-only model, therefore it does not have a reciprocity parameter. In addition, it uses the conditional independence assumption according to which the conditional and the marginal distributions coincide. For this reason, the closed-form conditional and joint do not apply for this method.}
\centering
\begin{adjustbox}{angle=0}
\resizebox{1\textwidth}{!}{%
{\renewcommand{\arraystretch}{1.2}
\setlength\arrayrulewidth{0.7pt}
\begin{tabular}{{l}*{3}{c}}
\toprule
                                   & \jcrep                                               & \crep                           & \mt                     \\ \midrule
Networks                    & Binary                                             & Weighted                    & Weighted                                                   \\
Likelihood                   & Bivariate Bernoulli     			 & Poisson 	 	      & Poisson      \\
Marginal mean           &  $\Exp\rup{A_{ij}}= \f{\lambda_{ij}+\eta \lambda_{ij}\lambda_{ji}}{Z_{(ij)}}$     &  $\Exp\rup{A_{ij}}= \f{\lambda_{ij}+\eta \lambda_{ji}}{1-\eta^2}$   &    $\Exp\rup{A_{ij}}= \lambda_{ij}$    \\
Conditional mean       & $\Exp\rup{A_{ij}|A_{ji}} =\f{\eta^{A_{ji}}\lambda_{ij}}{\eta^{A_{ji}}\lambda_{ij}+1}$    &     $\Exp\rup{A_{ij}|A_{ji}} = \lambda_{ij}+\eta A_{ji}$        &   $\Exp\rup{A_{ij}|A_{ji}}  = \Exp\rup{A_{ij}}$    \\
Relationship $\eta$ \textit{vs} \rec                         & Sublinear     		& Linear                                              & --                                           \\
Contribution $\lambda$ \textit{vs} $\eta$                & Multiplicative              & Additive                                           & --                       \\
Contribution \rec          					& Real      & Non negative                            & --                        \\
Closed-form marginal          & Yes      & No                                                 & Yes                                                      \\
Closed-form conditional        & Yes     & Yes                                                & --                                                      \\
Closed-form joint                & Yes     & No                                                 & --                                                   \\
\bottomrule                                    
\end{tabular}}}
\end{adjustbox}
\end{table*}

\section{Results}
In this section, we present the results obtained in synthetic and real networks. For comparison we use \crep, the model that combines communities and reciprocity with a pseudo-likelihood approximation~\cite{safdari2020generative}, and \mbox{{\small \textsc{MultiTensor }}}(\mt), a community detection-only generative model with a maximum likelihood approach~\cite{debacco2017multitensor}. Even if both of them posit a Poisson likelihood, in this work we use only binary networks for fair comparisons with our model \jcrep. We summarize the main similarities and differences among the models used in the analysis in \Cref{tab:crepvsjcrep}.

\subsection{Results on synthetic data}
We first validate the performance of the different methods on synthetic data generated with the  model proposed in this work. Being a generative model, given as input an initial set of parameters, one can draw a directed network with a community structure and a reciprocity value from the expression in \Cref{eqn:initial_joint}. The generative process is described in detail in \Cref{app:generativemodel}. We analyse networks with $N=1000$ nodes,  $K=2$ overlapping communities, $\langle k \rangle = 20$ average degree and different values of the pair-interaction parameter $\eta$ such that we obtain networks with  reciprocity values ($\mathsf{r}$) in the interval $[0, 0.8]$. We generate $10$ random samples for each value of \rec. In addition to these results, we provide further details for synthetic networks generated with different values of average degree in \Cref{app:synAvgDegree}. We test the ability of the models to i) recover the communities, ii) perform edge prediction tasks, and iii) generate sample networks that replicate relevant network quantities. 

\subsubsection{Community detection} 
\begin{figure}[t]
    \centering
 \makebox{\includegraphics[width=0.7\linewidth]{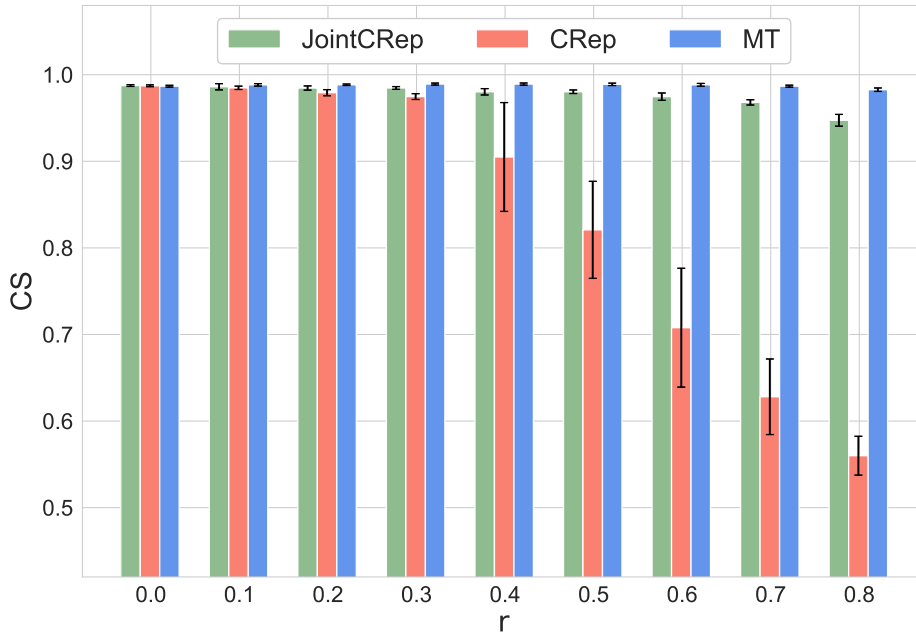}}
    \caption{\label{fig:cs_joints} \textbf{Community detection in synthetic networks.} Cosine similarity (CS) in synthetic networks with $N=1000$ nodes, $K=2$ overlapping communities, $\langle k \rangle = 20$ average degree, and different values of reciprocity \rec. Results are averages and standard deviations over $10$ synthetic networks.}
\end{figure}

To evaluate the performance of the methods on recovering the communities, we use the cosine similarity (CS), a measure useful to capture mixed-membership communities, as in this case. It ranges from 0 to 1, where 1 means perfect recovery. We calculate the average of the cosine similarities of both membership matrices $u$ and $v$, and then averaging over the nodes. The results are shown in \Cref{fig:cs_joints}. In the scenarios with low reciprocity values ($\mathsf{r} < 0.4$) all models perform well. However, as  \rec\text{} increases, \crep\text{} worsens while \jcrep\text{} keeps having good results comparable to those of the community-only algorithm, \mt. The big drop of \crep\text{} is due to the fact that this model gives increasingly less weight to communities as reciprocity increases, as pointed out in \citet{safdari2020generative}. Conversely, \jcrep\text{} is not affected by the different reciprocity values of the data and still performs as good as \mt\text{}, even by adding another parameter to the model. 

\subsubsection{Edge prediction}
The edge prediction task consists in estimating the existence of an edge by using the inferred parameters. The main quantity used as a score for the estimation of the entries of the adjacency matrix $A$ is the expected value of the marginal distribution. However, our model also provides the conditional distribution; hence its expected value can also be used as a score.  The difference lies in the nature of the question we try to answer. We use the marginal distribution to merely predict the existence of an edge, without considering additional information. On the other hand, with the conditional distribution, we  ask what is the probability of an edge $i \rightarrow j$, conditioned on observing the state of the opposite edge $j \rightarrow i$, i.e., knowing if it exists or not. Here, we exploit the presence or the absence of the edge in the opposite direction to better predict each given entry. Furthermore, our model specifies a joint distribution over the edges of a pair of nodes, and this allows us to answer questions more accurately compared to the standard models, which do not specify a joint distribution. For instance, what is the probability of \textit{jointly} observing both edges or even only an edge in one direction while not observing the other in the opposite? Our model directly captures this by specifying $P(A_{ij},A_{ji}| \boldsymbol \Theta)$, while others positing a conditional independence assumption can only compute an approximation as $P(A_{ij}| \boldsymbol \Theta) \,P(A_{ji}| \boldsymbol \Theta)$.

In our experiments below, we test edge prediction with various scores by using $5$-fold cross-validation. Specifically, we divide the dataset into five equal-size groups (folds) and train the models on four of them (training data) for learning the parameters; this contains 80\% of the possible pairs of nodes in the network, so that we hide pairs of entries $(A_{ij},A_{ji})$ from the training. One then predicts the existence of edges in the held-out group (test set). As performance metric, we measure the AUC on the test data, i.e., the probability that a randomly selected edge has higher expected value than a randomly selected non-existing edge. We compute both the regular, and conditional AUC values. To estimate the regular AUC, we take  the expected value $\Exp_{P(A_{ij} |\Theta)}[A_{ij}]$ as the score; while for the conditional AUC,  the expected value over the conditional distribution, i.e., $\Exp_{P(A_{ij} | A_{ij}, \Theta)}[A_{ij}]$ acts as the score. The latter cannot be computed for the community detection-only algorithm, as the marginal distribution is the same as the conditional, and thus the two AUC values coincide. We provide more details in \Cref{app:edgeprediction}, where we also show the ability of our model on edge prediction tasks by using the joint distribution.

\begin{figure}[H]
    \centering
    \makebox{\includegraphics[width=0.7\linewidth]{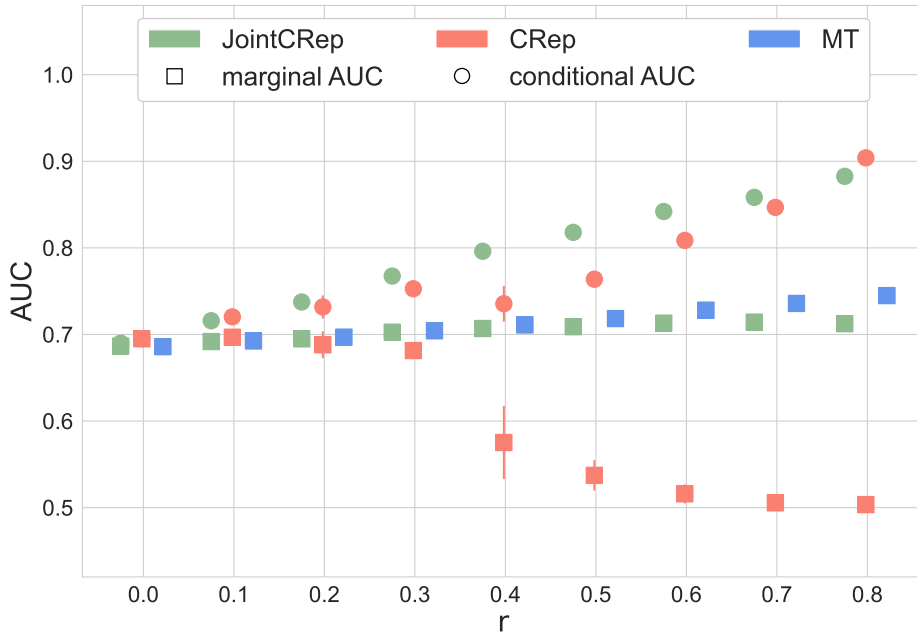}}
    \caption{\label{fig:auc_mar_cond}\textbf{Edge prediction in synthetic networks.} Synthetic networks with $N=1000$ nodes, $K=2$ overlapping communities, $\langle k \rangle = 20$ average degree, and different values of reciprocity \rec. Results are averages and standard deviations over $10$ synthetic networks and over $5$-folds of cross-validation test sets. Edge prediction performance is measured with AUC and the baseline is the random value $0.5$.}
\end{figure}

\Cref{fig:auc_mar_cond} displays the results of the marginal and conditional edge prediction for the different models. \jcrep\text{} significantly improves the performance of \crep\text{} when using the marginal expected value, and it performs as good as \mt. The latter, however, is not able to exploit the additional information given by the existence (or non-existence) of the edge in the opposite direction. This dependence is crucial in networks with reciprocity, i.e., most of the real world datasets, and models with an explicit conditional distribution can better adopt this information to obtain higher performance in edge prediction. Indeed, \jcrep\text{} and \crep\text{}  perform remarkably in this task, and our model presents more robust results both in terms of standard deviation, and growth. 

\subsubsection{Reproducing the topological properties}
\begin{figure}[t]
    \centering
    \makebox{\includegraphics[width=0.7\linewidth]{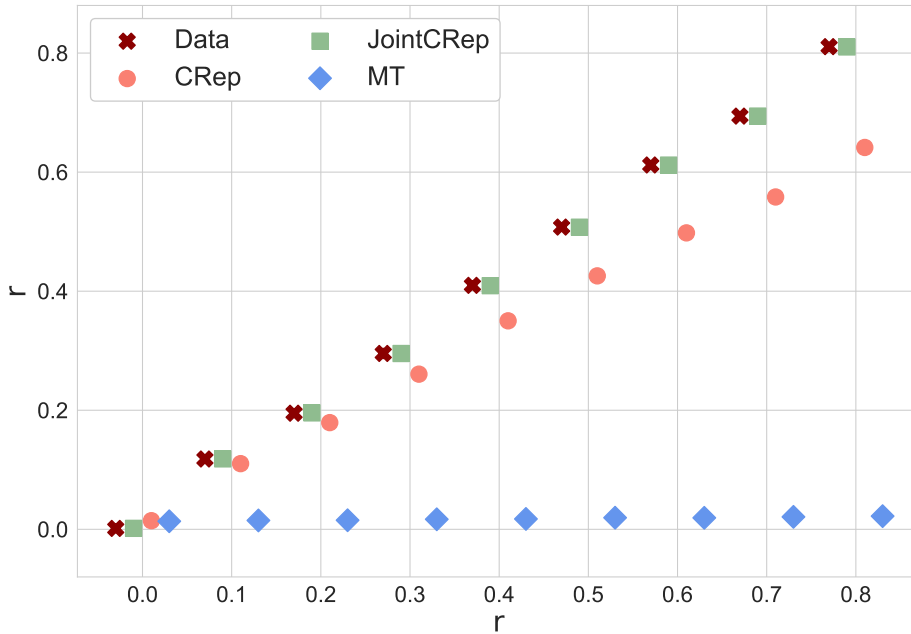}}
    \caption{\label{fig:sampled_r} \textbf{Reciprocity in sampled synthetic networks.}  Synthetic networks with $N=1000$ nodes, $K=2$ overlapping communities, $\langle k \rangle = 20$ average degree, and different values of reciprocity \rec. Results are empirical averages and standard deviations over $50$ samples of $10$ independent synthetic networks (five samples per input network). We measure the reciprocity and the dark red markers indicate the average on $10$ input networks.}
\end{figure}

A notable property of generative models is their ability to produce synthetic networks based on real-world datasets, such that the synthetic networks imitate the topological features of the real datasets. Following the approach in \citet{safdari2020generative}, for each individual network, we infer the network parameters by applying each model. Then, we use these inferred parameters to generate five network samples. We compare topological properties of these samples with those observed in the ground truth networks used to infer the parameters. In particular, we are interested in measuring reciprocity.  \Cref{fig:sampled_r} shows the performance of each model in reproducing this feature in sampled networks.  As it is expected, \mt\text{} is not capable of reflecting the observed value of the reciprocity in the ground truth network, a clear indication of the shortcoming of models based purely on community structure,  which indeed limits their applications.  Conversely, \jcrep\text{} perfectly reproduces this quantity. \crep\text{} generates sampled  networks with reciprocity lower than the ground truth due to the fact that it uses a Poisson likelihood resulting in weighted networks. Additional results are provided  in \Cref{app:propertysample}. \\

To summarize  the results on synthetic networks, \jcrep\text{} is capable of recovering communities on networks with varying reciprocity values, performing as good as models that are based purely on community structure. This capability overcomes the limitations of the recent \crep\text{} model. Moreover, \jcrep\text{} includes many performance enhancements in the edge prediction task, i.e.,  showing improved results in terms of marginal AUC and more robust conditional AUC values. Furthermore, \jcrep\text{} is also capable of generating  sampled networks with topological features that resemble that of the real data, e.g.,  reciprocity and average degree. Collectively, these findings show that \jcrep\text{} is able to overcome the limitations of both the community detection-only algorithm \mt\text{}  and the model that incorporates reciprocity through the pseudo-likelihood approximation \crep.

\subsection{Analysis of a high-school social network}\label{sec:highschool}
\begin{figure*}[ht]
    \makebox{\includegraphics[width=\textwidth]{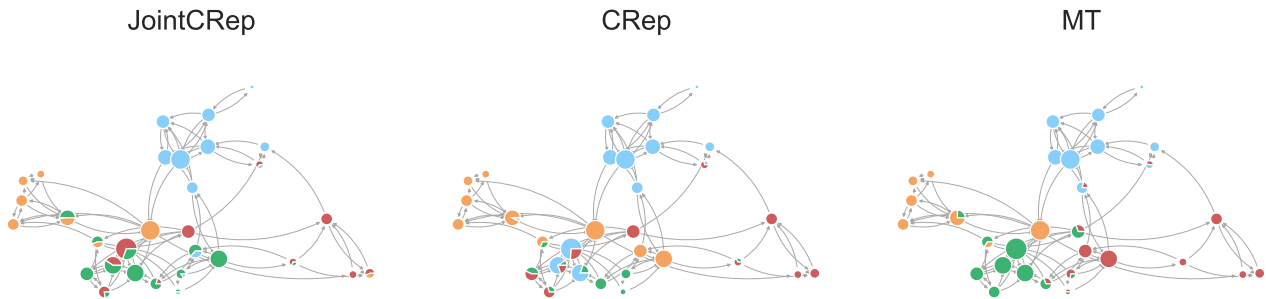}}
    \caption{\label{fig:HST11_community_soft_u} \textbf{Community detection in the high-school social network.} Mixed-membership partitions determined by the matrix $u$ inferred by \jcrep, \crep, and \mt.  Node size is proportional to the degree (in- and out-degree).}
\end{figure*}

We now study the social network that describes friendships between boys in a small high-school in Illinois that was collected in the fall of $1957$ \cite{konect:coleman}. Here, a node represents a boy and an edge from an individual $i$ to $j$ shows that node $i$ claimed to be friend of node $j$. We pre-processed the dataset by removing self-loops and isolated nodes. The resulting directed network has $31$ nodes, $100$ edges and reciprocity equal to $0.52$, i.e., only half of the edges (friendship relationships) are reciprocated. There is no additional metadata to describe the nodes, nor is there an available ground truth for the latent parameters. Therefore, we estimate the number of communities $K$ by performing edge prediction tasks via 5-fold cross-validation with different values of $K$. For each method the best performance in terms of AUC was achieved with $K=4$. Edge prediction also serves as model validation routine in the absence of ground truth information, as it is the case here. We found that results vary depending on the metric considered for evaluation, but in general they confirm that all models are fitting the data well, considering that the dataset is small and highly sparse, thus making prediction tasks hard. Further details for the edge prediction task are in \Cref{app:HST11_app}. 
\Cref{fig:HST11_community_soft_u} visualizes the mixed-membership partitions resulting from the matrix $u$, inferred by the different methods  (similar results are obtained for~$v$). Here we use the inferred value of $u$, which is  obtained from the run with the highest log-likelihood over $100$ random initializations of the parameters. All the algorithms assign most of the students to the same groups, except from a central block.  Here, \mt\text{} infers mostly hard memberships and balances the number of nodes in each cluster. Instead, \crep\text{} allocates only three nodes with small degree to the green community while places the nodes with higher degree in other clusters. \jcrep\text{}, shows a partition that lies in between, by inferring mixed-memberships for those nodes known as \textit{bridges}. To measure quantitatively the diversity of communities inferred by the various methods, we compute a modularity for directed networks and overlapping communities using different aggregation functions, as proposed by \citet{nicosia2009extending}.  Modularity assumes all communities to have statistically similar properties, in particular to have similar sizes, and it is suited for assortative community structure \cite{fortunato2007resolution,newman2016equivalence}. The communities shown in \Cref{fig:HST11_community_soft_u} reflect these properties to a certain extent, and we report the results in terms of  overlapping modularity in  \Cref{tab:Qnicosia}. We find modularity values between 0.48 and 0.75, depending on the aggregation function, with \jcrep\text{} and \mt\text{} achieving the highest values. 
 
\begin{figure*}[h]
    \centering
    \makebox{\includegraphics[width=\textwidth]{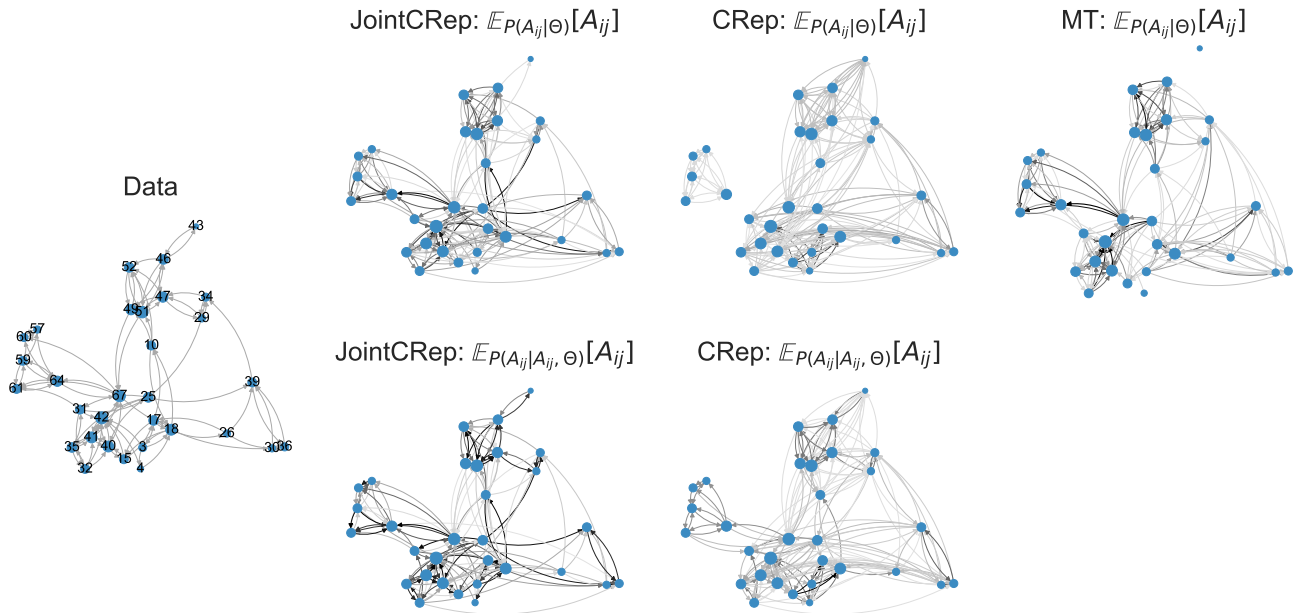}}
    \caption{\label{fig:HST11_graph} \textbf{High-school network reconstruction.} (left) High-school data and (right) network reconstructions by using as a score either the marginal expected value $\Exp_{P(A_{ij} | \Theta)}[A_{ij}]$ or the conditional expected value $\Exp_{P(A_{ij} | A_{ij}, \Theta)}[A_{ij}]$ with the inferred parameters. Note that the last is not available for \mt\text{} because the conditional and marginal distributions coincide. Edge width and darkness are proportional to the weight (given by the expected score); for visualization clarity we show only edges with weight greater than $0.2$. Node size is proportional to the degree (in- and out-degree) and node labels represent node IDs.}
\end{figure*}

Given the inferred parameters, we can test the ability of the models to reconstruct the input network, by using either the marginal expected value $\Exp_{P(A_{ij} | \Theta)}[A_{ij}]$, or the conditional expected value $\Exp_{P(A_{ij} | A_{ij}, \Theta)}[A_{ij}]$  as the  score. Note that the latter is not available for \mt\text{} because the conditional and marginal distributions coincide. \Cref{fig:HST11_graph} presents the results, where edge width and darkness of the reconstructed networks are proportional to the weight given by the expected score (for visualization clarity, we show only edges with weight greater than $0.2$). The network estimated by \crep\text{}, which  uses  the expected value of the marginals, does not capture the structure of the data magnificently, as it overestimates the presence of edges.  This model specifies conditional distributions and relies on a pseudo-likelihood approximation; since this approach is not necessarily accurate enough to approximate marginals, such results are expected.  Instead,  \mt\text{} and \jcrep\text{} estimate a sparser representation that is closer to the observed network. However, \mt\text{} is not able to notably  detect reciprocated edges, e.g., $(10,18)$ or $(64,67)$, while \jcrep\text{} remarkably recovers this type of interactions more precisely. For both \jcrep\text{} and \crep\text{}, including the conditional expected values improves their accuracy in reconstruction, resulting in identifying reciprocal edges correctly. The difference between the two models lies on the intensity: for instance \jcrep\text{} predicts the pair of edges between nodes $10$ and $18$  with a high probability, while \crep\text{} assigns a much lower probability to them. Hence, \jcrep\text{} is not only able to predict edges more precisely, but it also does so with higher probability. 
These qualitative observations are also confirmed by quantitative comparisons in terms of the Log Loss and the L1 Loss, two penalty metrics computed between the reconstructed and the true networks. They measure the difference between two input networks by taking into account the probability of the existence of an edge and computing a penalty for each mistake in predicting the observed value. A penalty of 0 denotes perfect reconstruction, as when assigning  probability 1 to the observed edge values. In general, lower values indicate higher similarity.  Further details can be found in \Cref{app:lossfunctions}.  We find that  \jcrep\text{} achieves the lowest values (i.e., better performance) in both metrics, with best overall performance achieved using the conditional expectation. See \Cref{tab:metricsReconstruction_HS} for details. 

\begin{figure*}[h]
    \centering
    \makebox{\includegraphics[width=\textwidth]{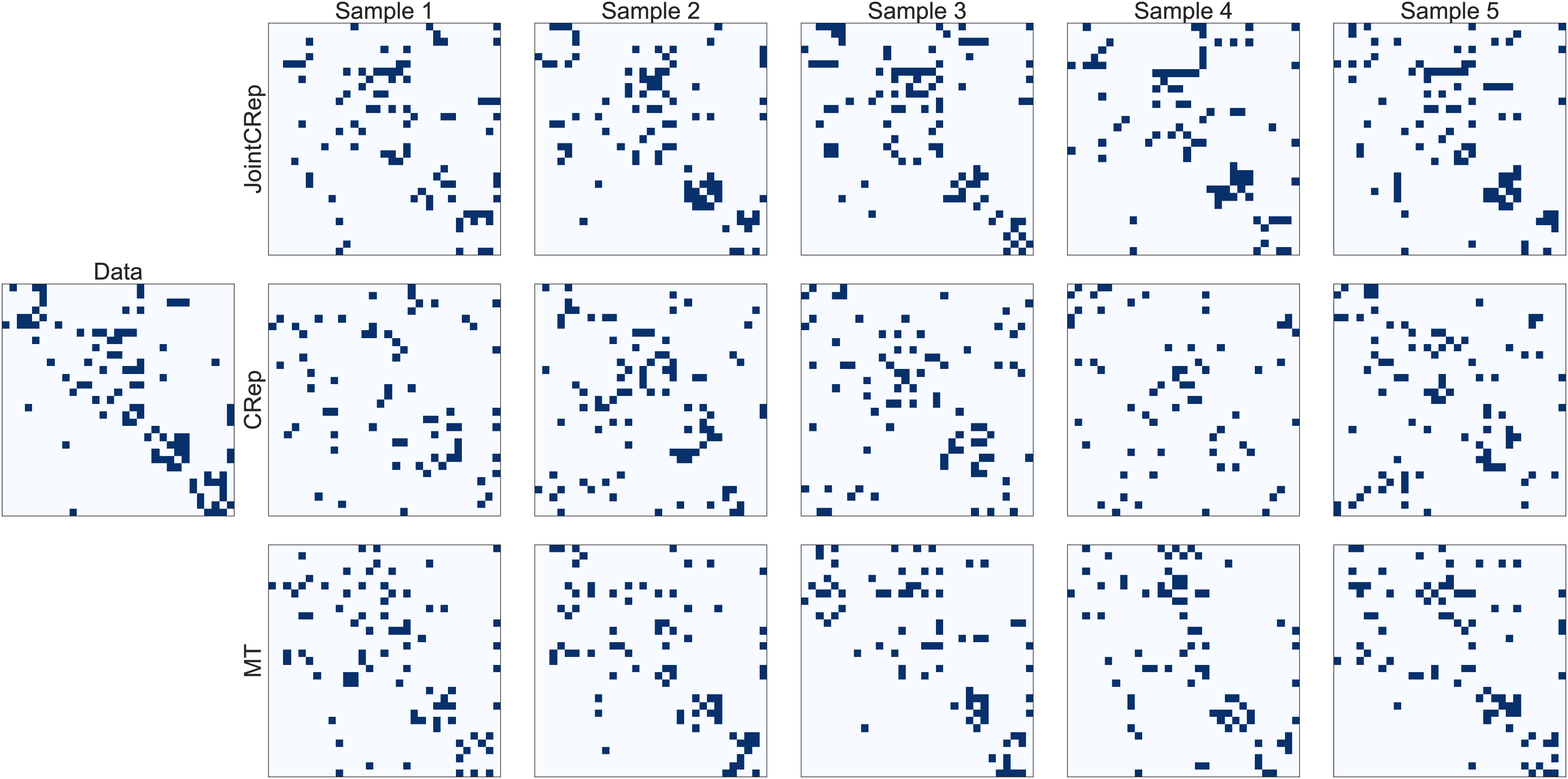}}
    \caption{\label{fig:HST11_adjacency_samples} \textbf{High-school network samples.} (left) High-school data and (right) five random  samples generated by  different methods with the inferred parameters. For each method, we first infer the parameters choosing the run with the highest log-likelihood over 100 random initializations of the parameters. Then, we generate the samples by using as input in the generative models the inferred parameters and the average degree of the original data. The generative process of \jcrep\text{} is described in \Cref{app:generativemodel}; for \crep\text{} we use the formulation described in \cite{safdari2020generative}; \mt\text{} follows the formulation of a standard mixed-membership variant of a stochastic block model, as described in \cite{debacco2017multitensor}.}
\end{figure*}

To further compare the strength of these methods, we examine their performance in generating samples that  resemble the observed network. For each model, we use the inferred parameters to generate five synthetic networks, as shown in \Cref{fig:HST11_adjacency_samples}. Again, we notice how the samples generated by \jcrep\text{} better resemble the observed network, as it is easier to distinguish the four blocks generated by \jcrep \text{}, compared to the  samples from the other algorithms. In particular, \jcrep \text{} finds denser groups given by reciprocated edges. In addition to these qualitative results, \Cref{tab:high_school_samples} reports the topological properties of the observed data and the sampled networks, showing that \jcrep\text{} generates networks samples that on average are most similar to the observed data in terms of average degree, reciprocity and clustering coefficient.

\subsection{Analysis of a vampire bat network}
\begin{figure*}[h]
    \centering
    \makebox{\includegraphics[width=\textwidth]{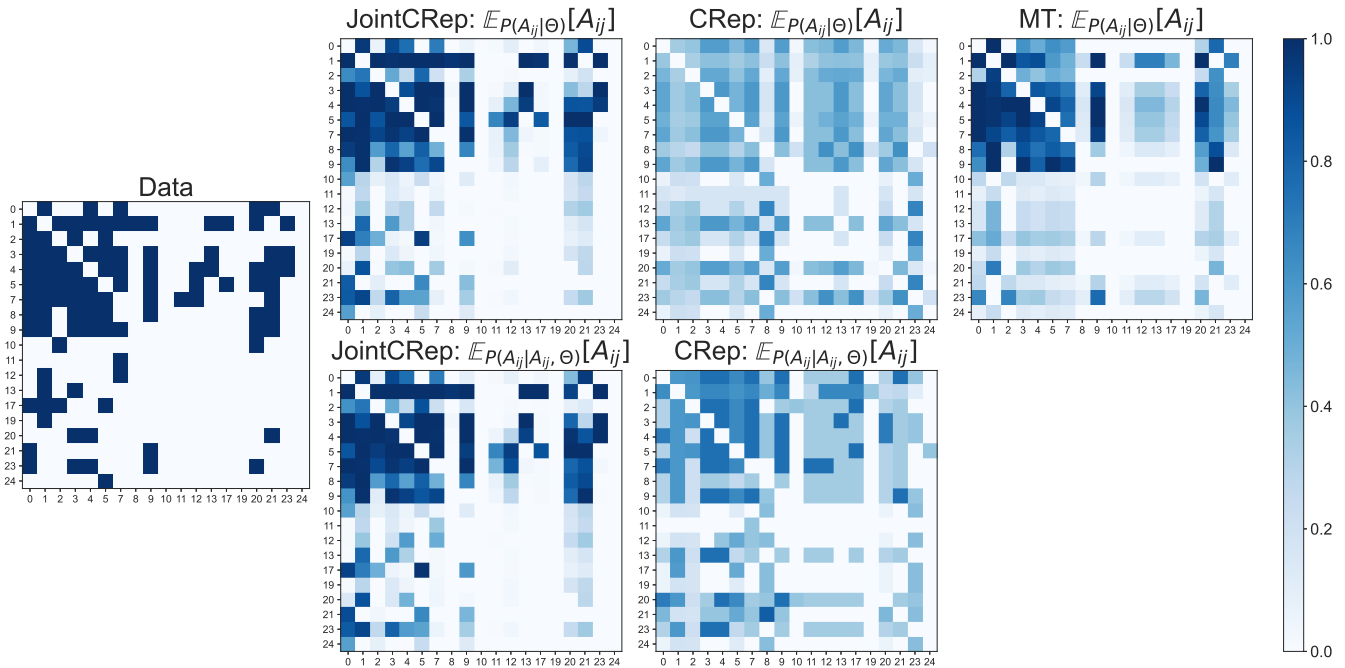}}
    \caption{\label{fig:vampire_bat_adjacency} \textbf{Vampire bat network reconstruction.} (Left) The adjacency matrix of the vampire bat data and (right) its estimates by using as a score either the marginal expected value $\Exp_{P(A_{ij} | \Theta)}[A_{ij}]$ or the conditional expected value $\Exp_{P(A_{ij} | A_{ij}, \Theta)}[A_{ij}]$ with the inferred parameters. Note that the last is not available for \mt\text{} because the conditional and marginal distributions coincide. The intensity of the entries is proportional to the score probability, as shown in the colorbar. The labels near the ticks represent node IDs.}
\end{figure*}

As a second example, we study the network of food sharing interactions in captive vampire bats,  collected by \citet{carter2013food}. These animals often regurgitate food to roost-mates that fail to feed. The decision of who to feed may depend on both kin relatedness and reciprocal sharing. Hence, in this dataset we expect reciprocity to be an important factor for tie formation. In the study, they fasted 20 vampire bats and induced food sharing on 48 days, over a 2 year period. They showed that reciprocal sharing predicts future food regurgitation more than relatedness or other non-kin food sharing behaviors, such as harassment. From the collected data,  we construct  a directed network by building an edge from a bat $i$ to another $j$ if node $i$ fed  $j$ at least once. We removed isolated nodes and obtained a network with $19$ nodes, $103$ edges and reciprocity equal to $0.64$. We fix the number of communities $K=2$ and analyse the data with the different methods. As for the high-school data, results of edge prediction tasks for model validation confirm that all the models represent the data well, see  \Cref{tab:CV_vampire} for details. We are now interested in measuring the ability of the models to recover the observed network with the inferred parameters, in particular their ability to recover topological properties  such as reciprocity. To this aim, we consider the marginal and the conditional expected values, as in \Cref{sec:highschool}. \Cref{fig:vampire_bat_adjacency} shows the adjacency matrix of the data and its different estimates, obtained by each method. The network embodies a core-periphery structure, where the main core (labels $0-9$) is made of female bats. \jcrep\text{} recovers this structure  more accurately than other methods, the off-diagonal entries show this fascinating result clearly, while the other methods overestimate the amount of edges. Similarly as observed in the high-school network, our model is not only more accurate, but also assigns higher probabilities to these entries and best performs both in terms of Log and L1 Loss, see \Cref{tab:metricsReconstruction_vampire}.

\begin{figure}[!h]
    \centering
    \makebox{\includegraphics[width=0.7\linewidth]{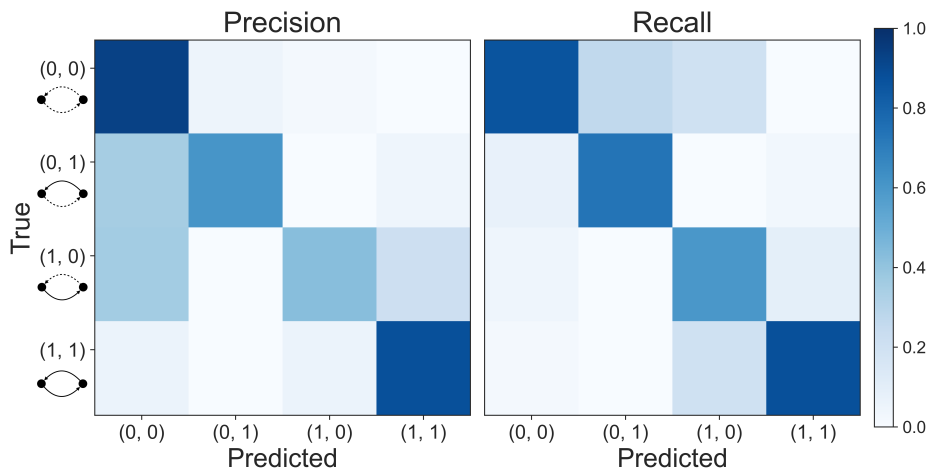}}
    \caption{\label{fig:vampire_bat_confmatr} \textbf{Precision and recall of the vampire bat network.} Statistics based on the confusion matrix that compares the entries of the adjacency matrix and the estimates obtained with the joint distribution of \jcrep. The precision is given by a normalization by row, while the recall accounts for the normalization by column. The label $(0,0)$ denotes no interactions between nodes $i$ and $j$; labels $(0,1)$ and $(1,0)$ considers the pair of edges where only one edge in one direction is present, and the label $(1,1)$ indicates  reciprocated edges.}
\end{figure}

In addition to the marginal and conditional expected value, we can consider the joint distribution to estimate the entries of the adjacency matrix. This is equivalent to assign a value to each pair $(A_{ij} , A_{ji})$ from the set $\{(0,0),(0,1),(1,0),(1,1)\}$, that transforms the edge prediction task into a classification problem. We predict the label associated to the highest probability among $[p_{00}, p_{01}, p_{10}, p_{11}]$, where these are computed by using \Crefrange{eqn:p01}{eqn:p00} with the inferred parameters. We assess the goodness of our performance by computing the precision and recall of the predicted labels versus the true labels, as shown in \Cref{fig:vampire_bat_confmatr}. The precision identifies the proportion of correctly classified observed entries. The figure illustrates high precision values consistently across edge labels, as the highest entries are along the diagonal. In particular, \jcrep\text{} is able to correctly classify  the pairs $(0,0)$ and $(1,1)$. Observing where our model misclassifies, this mainly happens by predicting no edges, i.e., assign label $(0,0)$, when the true ones are either $(0,1)$ or $(1,0)$,  implying a tendency to  estimate sparser networks. On the other hand, the recall indicates the proportion of predicted edges being correctly classified. Also in this case, the highest entries are in the main diagonal and in predicting  the pairs $(0,0)$ and $(1,1)$. Overall, in this case we obtain higher intensities as for the precision, indicating the tendency of labeling the predicted edges with the right type. 

To conclude our analysis, we compare five random samples generated with the inferred parameters of each model and check whether they reproduce topological properties as those observed in the real data.  \Cref{tab:vampire_bat_samples} shows that \jcrep\text{} outperforms  other models in terms of all topological properties. In particular, it generates sampled networks with reciprocity values closest to the real data, but also reproduces realistic values of the clustering coefficient.

\begin{table*}
\caption{\label{tab:vampire_bat_samples} \textbf{Topological properties in vampire bat and its sampled networks.} Results are averages and standard deviations over five samples. We measure the number of nodes $N$, the number of edges $M$, the average degree $\langle k \rangle$, the reciprocity \rec, and the clustering coefficient $cc$.}
\centering
\setlength{\tabcolsep}{10pt}
\begin{adjustbox}{angle=0}
\resizebox{0.9\linewidth}{!}{%
{\renewcommand{\arraystretch}{1.2}
\setlength\arrayrulewidth{0.7pt}
\begin{tabular}{*{6}{c}}
\toprule
                      & $N$    & $M$   &  $\langle k \rangle$  & \rec & $cc$                                        \\ \midrule
 Data   & 19                     & 103                     & 10.84                  & 0.64                  & 0.54 \\
 \jcrep & $ 18.4 \pm 0.89  $ & $ 100.4 \pm 5.41$   & $ 10.92 \pm 0.38 $ & $0.61 \pm 0.03 $ & $0.55 \pm 0.05$ \\
 \crep  & $ 18.2 \pm 0.84  $ & $ 74.2   \pm 5.40  $ & $8.16   \pm 0.54 $ & $0.51 \pm 0.04$  & $0.27 \pm 0.05 $ \\
 \mt     & $17.4 \pm 1.14  $ & $ 70.0      \pm 7.38  $ & $8.06   \pm 0.83 $ & $ 0.36 \pm 0.06 $ & $0.37 \pm 0.01$ \\           
\bottomrule                                    
\end{tabular}}}
\end{adjustbox}
\end{table*}

\section{Discussion and conclusion}
In this paper, we have presented a generative model called \jcrep\text{} that takes into account community structure and reciprocity by specifying a closed-form joint distribution of a pair of network edges, without relying upon approximations.
 Our model also provides closed-form analytical expressions for both the marginal and conditional distributions, and enables practitioners to address with more accuracy questions that were not fully captured by standard models; for instance, predicting the joint existence of mutual ties between pairs of nodes. 

We first validated our model by applying it to synthetic network datasets, where we achieved remarkable performance in recovering communities, edge prediction tasks, and generating synthetic networks that replicate topological features observed in real networks.  We then analyzed two real datasets that are relevant for social scientists and behavioral ecologists, where we found that \jcrep\text{} obtains more robust and interpretable results.
The results shown in this work highlight main benefits of using a model that considers closed-form joint distributions of pairs of edges in networks, while also showing possible shortcomings of other approaches. While it is difficult to pinpoint theoretical reasons for these shortcomings, the variety of experiments that we discussed throughout this manuscript show possible practical consequences of them. In particular, standard generative models make strong conditional independence assumptions that reflect in poor recovery of topological properties as reciprocity. On the other hand, models that specify only conditional distributions rely on pseudo-likelihood approximations that may reflect in weak recovery of latent parameters as communities in certain regimes.
 Collectively, our model is able to overcome the limitations of both these approaches thanks to the modeling of closed-form joint distributions while also keeping computational complexity under control.

The framework we described could be extended in a number of ways. \jcrep\text{} analyses binary and single-layer networks; therefore,  possible extensions could account for weighted and possibly multilayer networks,  where we have edges of different types. Another approach could consider dynamic networks, which have evolving structure over time, and adapt the parameters accordingly \cite{safdari2021reciprocity}. Moreover, our model captures the reciprocity through a unique pair-interaction parameter for the whole network. This model could be improved in the future by including node-dependent parameters in scenarios where reciprocity varies between individuals. Furthermore, many real world datasets contain attributes that provide additional information about their features. Incorporating  these extra informations on nodes could result in a more realistic analysis \cite{contisciani2020community}.

\jcrep\text{}, to the best of our knowledge, is the first such method for fully capturing reciprocity by jointly modeling pairs of edges with exact 2-edge joint distributions. We believe it will serve as a baseline for future models that tackle more complicated interactions that go beyond pairwise interaction, e.g., triadic closure \cite{peixoto2022disentangling}.

\section*{Acknowledgements}
\vspace{-0.1in}
The authors thank the International Max Planck Research School for Intelligent Systems (IMPRS-IS) for supporting Martina Contisciani. 
\textbf{Funding:} All the authors were supported by the Cyber Valley Research Fund. 
\textbf{Competing interests}: The authors declare that they have no competing interests.
 \section*{Data and materials availability}
\vspace{-0.1in}
An open-source algorithmic implementation of the model together with the code to generate synthetic data is publicly available and can be found at \href{https://github.com/mcontisc/JointCRep}{https://github.com/mcontisc/JointCRep}.

\bibliographystyle{apalike}
\bibliography{bibliography}

\begin{thebibliography}{}

\bibitem[Ball et~al., 2011]{ball2011efficient}
Ball, B., Karrer, B., and Newman, M.~E. (2011).
\newblock Efficient and principled method for detecting communities in
  networks.
\newblock {\em Physical Review E}, 84(3):036103.

\bibitem[Carter and Wilkinson, 2013]{carter2013food}
Carter, G.~G. and Wilkinson, G.~S. (2013).
\newblock Food sharing in vampire bats: reciprocal help predicts donations more
  than relatedness or harassment.
\newblock {\em Proceedings of the Royal Society B: Biological Sciences},
  280(1753):20122573.

\bibitem[Coleman, 1964]{konect:coleman}
Coleman, J.~S. (1964).
\newblock Introduction to mathematical sociology.
\newblock {\em London Free Press Glencoe}.

\bibitem[Contisciani et~al., 2020]{contisciani2020community}
Contisciani, M., Power, E.~A., and De~Bacco, C. (2020).
\newblock Community detection with node attributes in multilayer networks.
\newblock {\em Scientific reports}, 10(1):1--16.

\bibitem[Dai et~al., 2013]{dai2013multivariate}
Dai, B., Ding, S., Wahba, G., et~al. (2013).
\newblock Multivariate bernoulli distribution.
\newblock {\em Bernoulli}, 19(4):1465--1483.

\bibitem[De~Bacco et~al., 2021]{de2021latent}
De~Bacco, C., Contisciani, M., Cardoso-Silva, J., Safdari, H., Baptista, D.,
  Sweet, T., Young, J.-G., Koster, J., Ross, C.~T., McElreath, R., et~al.
  (2021).
\newblock Latent network models to account for noisy, multiply-reported social
  network data.
\newblock {\em arXiv preprint arXiv:2112.11396}.

\bibitem[De~Bacco et~al., 2017]{debacco2017multitensor}
De~Bacco, C., Power, E.~A., Larremore, D.~B., and Moore, C. (2017).
\newblock Community detection, link prediction, and layer interdependence in
  multilayer networks.
\newblock {\em Phys. Rev. E}, 95:042317.

\bibitem[Fell and Wagner, 2000]{fell2000small}
Fell, D.~A. and Wagner, A. (2000).
\newblock The small world of metabolism.
\newblock {\em Nature biotechnology}, 18(11):1121--1122.

\bibitem[Fortunato, 2010]{fortunato2010community}
Fortunato, S. (2010).
\newblock Community detection in graphs.
\newblock {\em Physics reports}, 486(3-5):75--174.

\bibitem[Fortunato and Barthelemy, 2007]{fortunato2007resolution}
Fortunato, S. and Barthelemy, M. (2007).
\newblock Resolution limit in community detection.
\newblock {\em Proceedings of the national academy of sciences}, 104(1):36--41.

\bibitem[Garlaschelli and Loffredo, 2004]{garlaschelli2004patterns}
Garlaschelli, D. and Loffredo, M.~I. (2004).
\newblock Patterns of link reciprocity in directed networks.
\newblock {\em Physical review letters}, 93(26):268701.

\bibitem[Goldenberg et~al., 2010]{goldenberg2010}
Goldenberg, A., Zheng, A.~X., Fienberg, S.~E., and Airoldi, E.~M. (2010).
\newblock A survey of statistical network models.
\newblock {\em Foundations and Trends in Machine Learning}, 2(2):129--233.

\bibitem[Hanley and McNeil, 1982]{hanley1982meaning}
Hanley, J.~A. and McNeil, B.~J. (1982).
\newblock The meaning and use of the area under a receiver operating
  characteristic (roc) curve.
\newblock {\em Radiology}, 143(1):29--36.

\bibitem[Holland et~al., 1983]{holland1983stochastic}
Holland, P.~W., Laskey, K.~B., and Leinhardt, S. (1983).
\newblock Stochastic blockmodels: First steps.
\newblock {\em Social networks}, 5(2):109--137.

\bibitem[Holland and Leinhardt, 1981]{holland1981exponential}
Holland, P.~W. and Leinhardt, S. (1981).
\newblock An exponential family of probability distributions for directed
  graphs.
\newblock {\em Journal of the american Statistical association},
  76(373):33--50.

\bibitem[Ising, 1925]{ising1925beitrag}
Ising, E. (1925).
\newblock Beitrag zur theorie des ferromagnetismus.
\newblock {\em Zeitschrift f{\"u}r Physik}, 31(1):253--258.

\bibitem[Karrer and Newman, 2011]{karrer2011stochastic}
Karrer, B. and Newman, M.~E. (2011).
\newblock Stochastic blockmodels and community structure in networks.
\newblock {\em Physical review E}, 83(1):016107.

\bibitem[Li et~al., 2019]{li2019reciprocity}
Li, W., Aste, T., Caccioli, F., and Livan, G. (2019).
\newblock Reciprocity and impact in academic careers.
\newblock {\em EPJ Data Science}, 8(1):20.

\bibitem[Newman, 2001]{newman2001structure}
Newman, M.~E. (2001).
\newblock The structure of scientific collaboration networks.
\newblock {\em Proceedings of the national academy of sciences},
  98(2):404--409.

\bibitem[Newman, 2016]{newman2016equivalence}
Newman, M.~E. (2016).
\newblock Equivalence between modularity optimization and maximum likelihood
  methods for community detection.
\newblock {\em Physical Review E}, 94(5):052315.

\bibitem[Newman et~al., 2002]{newman2002email}
Newman, M.~E., Forrest, S., and Balthrop, J. (2002).
\newblock Email networks and the spread of computer viruses.
\newblock {\em Physical Review E}, 66(3):035101.

\bibitem[Nicosia et~al., 2009]{nicosia2009extending}
Nicosia, V., Mangioni, G., Carchiolo, V., and Malgeri, M. (2009).
\newblock Extending the definition of modularity to directed graphs with
  overlapping communities.
\newblock {\em Journal of Statistical Mechanics: Theory and Experiment},
  2009(03):P03024.

\bibitem[Park and Newman, 2004]{park2004statistical}
Park, J. and Newman, M.~E. (2004).
\newblock Statistical mechanics of networks.
\newblock {\em Physical Review E}, 70(6):066117.

\bibitem[Peixoto, 2022]{peixoto2022disentangling}
Peixoto, T.~P. (2022).
\newblock Disentangling homophily, community structure, and triadic closure in
  networks.
\newblock {\em Physical Review X}, 12(1):011004.

\bibitem[Ready and Power, 2021]{ready_power_2021}
Ready, E. and Power, E.~A. (2021).
\newblock Measuring reciprocity: Double sampling, concordance, and network
  construction.
\newblock {\em Network Science}, page 1–16.

\bibitem[Robins et~al., 2007]{robins2007introduction}
Robins, G., Pattison, P., Kalish, Y., and Lusher, D. (2007).
\newblock An introduction to exponential random graph (p*) models for social
  networks.
\newblock {\em Social networks}, 29(2):173--191.

\bibitem[Safdari et~al., 2021]{safdari2020generative}
Safdari, H., Contisciani, M., and {De Bacco}, C. (2021).
\newblock Generative model for reciprocity and community detection in networks.
\newblock {\em Phys. Rev. Research}, 3:023209.

\bibitem[Safdari et~al., 2022]{safdari2021reciprocity}
Safdari, H., Contisciani, M., and De~Bacco, C. (2022).
\newblock Reciprocity, community detection, and link prediction in dynamic
  networks.
\newblock {\em Journal of Physics: Complexity}.

\bibitem[Seshadhri et~al., 2020]{seshadhri2020impossibility}
Seshadhri, C., Sharma, A., Stolman, A., and Goel, A. (2020).
\newblock The impossibility of low-rank representations for triangle-rich
  complex networks.
\newblock {\em Proceedings of the National Academy of Sciences},
  117(11):5631--5637.

\bibitem[Snijders et~al., 2006]{snijders2006new}
Snijders, T.~A., Pattison, P.~E., Robins, G.~L., and Handcock, M.~S. (2006).
\newblock New specifications for exponential random graph models.
\newblock {\em Sociological methodology}, 36(1):99--153.

\bibitem[Wasserman and Anderson, 1987]{wasserman1987stochastic}
Wasserman, S. and Anderson, C. (1987).
\newblock Stochastic a posteriori blockmodels: Construction and assessment.
\newblock {\em Social networks}, 9(1):1--36.

\bibitem[Wasserman et~al., 1994]{wasserman1994social}
Wasserman, S., Faust, K., et~al. (1994).
\newblock {\em Social network analysis: Methods and applications}.
\newblock Cambridge university press.

\bibitem[Watts and Strogatz, 1998]{watts1998collective}
Watts, D.~J. and Strogatz, S.~H. (1998).
\newblock Collective dynamics of ‘small-world’ networks.
\newblock {\em Nature}, 393(6684):440--442.

\bibitem[Williams and Martinez, 2000]{williams2000simple}
Williams, R.~J. and Martinez, N.~D. (2000).
\newblock Simple rules yield complex food webs.
\newblock {\em Nature}, 404(6774):180--183.

\end{thebibliography}

\newcommand{\beginsupplement}{%
        \setcounter{table}{0}
        \renewcommand{\thetable}{S\arabic{table}}%
        \setcounter{figure}{0}
        \renewcommand{\thefigure}{S\arabic{figure}}%
        \setcounter{equation}{0}
        \renewcommand{\theequation}{S\arabic{equation}}
         \setcounter{section}{0}
        \renewcommand{\thesection}{S\arabic{section}}
 }

\clearpage
\beginsupplement
\begin{widetext}

\appendix
\section*{{Supporting Information (SI)}}

\section{Detailed derivations} \label{app:derivations}
Combining \Crefrange{eqn:ftot}{eqn:Jij} we get the explicit mapping between the latent variables and the instances of the joint probability in \Cref{eqn:initial_joint}:
\begin{align}
p_{01} &= \f{\lambda_{ji}}{Z_{(ij)}} \label{eqn:p01} \\
p_{10} &= \f{\lambda_{ij}}{Z_{(ij)}} \label{eqn:p10} \\
p_{11} &= \f{\eta\lambda_{ij}\lambda_{ji}}{Z_{(ij)}} \label{eqn:p11} \\
p_{00} &= \f{1}{Z_{(ij)}} \label{eqn:p00} \ ,
\end{align}
where the normalization constant is:
\be\label{eqn:Z}
Z_{(ij)} = \lambda_{ij}+\lambda_{ji}+\eta  \lambda_{ij}\lambda_{ji}+1 \ .
\ee
One property of the bivariate Bernoulli is that both marginal and conditional distributions are univariate Bernoulli. Thus, the marginal distributions of $A_{ij}$ and $A_{ji}$ have densities:
\begin{align}\label{eqn:marginals}
P(A_{ij}) &= (p_{10} + p_{11})^{A_{ij}}(p_{00} + p_{01})^{(1-A_{ij})}\\
P(A_{ji}) &= (p_{01} + p_{11})^{A_{ji}}(p_{00} + p_{10})^{(1-A_{ji})} \ ,
\end{align}
while the conditional distributions are the following:
\begin{align}\label{eqn:conditionals}
P(A_{ij}| A_{ji}) &=\bup{\f{p(1, A_{ji})}{p(1,A_{ji})+p(0,A_{ji})}}^{A_{ij}}\bup{ \f{p(0, A_{ji})}{p(1,A_{ji})+p(0,A_{ji})}}^{(1-A_{ij})}\\
P(A_{ji}| A_{ij}) &=\bup{ \f{p(A_{ij}, 1)}{p(A_{ij},1)+p(A_{ij},0)}}^{A_{ji}}\bup{ \f{p(A_{ij}, 0)}{p(A_{ij},1)+p(A_{ij},0)}}^{(1-A_{ji})} \quad .
\end{align}
In addition to the expected values reported in the manuscript, we can also compute the variances and the covariance between the random variables:
\begin{align}
\Var\rup{A_{ij}} &=\bup{\f{\lambda_{ij}(1+\eta\lambda_{ji})}{Z_{(ij)}}}\bup{\f{1+\lambda_{ji}}{Z_{(ij)}}} \\
\Var\rup{A_{ji}} &=\bup{\f{\lambda_{ji}(1+\eta\lambda_{ij})}{Z_{(ij)}}}\bup{\f{1+\lambda_{ij}}{Z_{(ij)}}} \\
\Cov\rup{A_{ij},A_{ji}} &=\f{\eta\lambda_{ij}\lambda_{ij}-\lambda_{ij}\lambda_{ij}}{Z_{(ij)}^2}\ .
\end{align}
At each step of the EM algorithm one updates $\rho$ using \Cref{eqn:rho} (E-step) and then maximizes $\mathcal{L}(\rho,\boldsymbol \Theta)$ with respect to $\boldsymbol \Theta=(u, v, w, \eta)$ by setting partial derivatives to zero (M-step). The derivative w.r.t. $\eta$ is given by:
\begin{align} \label{eqn:der_eta}
\f{\partial \mathcal{L}(\rho,\boldsymbol \Theta)}{\partial \eta} =  \f{1}{2\eta}\sum_{i,j}A_{ij}A_{ji} - \f{1}{2}\sum_{i,j}\f{\lambda_{ij}\lambda_{ji}}{\lambda_{ij}+\lambda_{ji}+\eta  \lambda_{ij}\lambda_{ji}+1}\overset{!}{=}0 \quad,
\end{align}
that leads to:
\be \label{eqn:update_eta}
\eta = \f{\sum_{i,j}A_{ij}A_{ji}}{\sum_{i,j}\rup{\f{\lambda_{ij}\lambda_{ji}}{\lambda_{ij}+\lambda_{ji}+\eta  \lambda_{ij}\lambda_{ji}+1}}} \quad.
\ee
Similarly, we get the updates for $u, v$ and $w$:
\begin{align} 
u_{ik} &=\f{ \sum_{j,q}A_{ij}\rho_{ijkq}}{\sum_{j}\rup{\f{\sum_{q}v_{jq}w_{kq}(1+\eta \lambda_{ji})}{\lambda_{ij}+\lambda_{ji}+\eta \lambda_{ij}\lambda_{ji}+1}}}  \\ 
v_{ik} &=\f{ \sum_{j,q}A_{ji}\rho_{jiqk}}{\sum_{j}\rup{\f{\sum_{q}u_{jq}w_{qk}(1+\eta \lambda_{ij})}{\lambda_{ij}+\lambda_{ji}+\eta \lambda_{ij}\lambda_{ji}+1}}} \\ 
w_{kq} &=\f{ \sum_{i,j}A_{ij}\rho_{ijkq}}{\sum_{i,j}\rup{\f{u_{ik}v_{jq}(1+\eta \lambda_{ji})}{\lambda_{ij}+\lambda_{ji}+\eta \lambda_{ij}\lambda_{ji}+1}}} \ .
\end{align}

\section{Benchmark generative model}\label{app:generativemodel}
The model we propose in the manuscript is able to generate synthetic data with intrinsic community structure and a reciprocity value. It takes as input a set of membership vectors, $\boldsymbol{u_i}$ and $\boldsymbol{v_i}$, affinity matrix $w$, and a pair-interaction parameter $\eta$; the output is a directed and binary network with adjacency matrix $A$ whose pairs of edges are conditionally independent from each other. We use the same formulation as in \citet{safdari2020generative}, but our approach differs in that edges between a given pair of nodes are generated stochastically according to the joint probability in \Cref{eqn:initial_joint}, and not according to a two-step sampling procedure. In detail, we assign a value to each pair $(A_{ij},A_{ji})$ by considering the vector of cumulative probabilities built using \Crefrange{eqn:p01}{eqn:p00}. To enforce sparsity, we multiply $\lambda$ by a constant $\zeta$, and in order to automatically rescale the expected value in \Cref{eqn:marginalmean} we have to impose 
\begin{align}
\Exp\rup{M} =  \sum_{i,j}\f{\zeta \, \lambda_{ij}+\eta \, \zeta \, \lambda_{ij}\, \zeta\, \lambda_{ji}}{\zeta\,\lambda_{ij}+\zeta\,\lambda_{ji}+\eta \,\zeta\, \lambda_{ij}\,\zeta\,\lambda_{ji}+1} \label{eqn:constant}
\end{align}
and solve with respect to $\zeta$, where $\Exp\rup{M}$ is the expected number of edges, a quantity given in input. The benchmark we propose here differs from the one presented in \citet{safdari2020generative} for multiple reasons, as we summarize in \Cref{tab:crepvsjcrep}. In addition to those, it is worth mentioning that the competing benchmark in \citet{safdari2020generative} depends on a variable,  $cr_{ratio} = 1 - \eta$, that controls the proportion of edges generated purely by either community or reciprocity effect. This implies that in order to have high reciprocity we may weaken the impact of community effect. This does not happen with the model we propose here, as tie formation can be highly influenced by both reciprocity and community structure at the same time, thus providing a more reliable and truthful representation in certain real world examples. 

In the manuscript, we use networks generated with the benchmark proposed above where we fix $N = 1000$ nodes, $K = 2$ overlapping communities,  $\langle k \rangle = 20$ average degree, and different values of the pair-interaction parameter $\eta$ such that we obtain networks with reciprocity values \rec\text{} in the interval $[0, 0.8]$. In detail, we use $\eta \in \{0.1, 10, 20, 40, 80, 140, 280, 500, 1500\}$ to get \rec\text{} $\in \{0, 0.1, 0.2, \dots, 0.8\}$, and we generate 10 different samples for each value of $\eta$. Additionally to the data presented in the manuscript, we also report in \Cref{app:synAvgDegree} further results on synthetic data generated by varying the average degree $\langle k \rangle$ in the interval $[2, 18]$ while fixing $\eta = 1000$ and the other parameters as above. To generate the membership matrices $u$ and $v$ we first assign an equal-size unmixed group membership and then we apply the overlapping to $20\%$ of the nodes by drawing those entries from a Dirichlet distribution with parameter $\alpha=0.1$. The affinity matrix $w$ is generated using an assortative block structure with main probabilities $p_1 = \langle k \rangle \, K \, / \, N$ and secondary probabilities $p_2 = 0.1 \, p_1$. Thus the latent variables $\boldsymbol \Theta = (u, v, w, \eta)$ are fixed. Then, edges are drawn according to the generative model described above.  

For sake of completeness, we also analysed synthetic networks generated with the model proposed in \citet{safdari2020generative} obtaining similar results and same conclusions. Furthermore, we investigated the behaviour of the models on networks with more than 2 communities and noticed that results are not highly impacted by this parameter. We do not report them here for sake of brevity. 

\section{Results on synthetic data}
\subsection{Edge prediction}\label{app:edgeprediction}
We test edge prediction by using a 5-fold cross-validation routine: we divide the dataset into five equal-size groups and train the model on four of them (training set) to infer the parameters; the fifth group is then used as test set to evaluate the existence of edges $A_{ij}$ (in this set). By varying which group we use as test set, we get five trials per realization and the final score is the average over these. To divide the dataset into five folds, we use a symmetric mask, i.e., in each trial the training set contains the 80\% of the possible entries $(A_{ij}, A_{ji})$.  In the manuscript we show the performance of the models in edge prediction when using the marginal and conditional expected values, $\Exp\rup{A_{ij}}$ and $\Exp\rup{A_{ij}|A_{ji}}$ respectively. Here, we measure the AUC that is equivalent to the area under the receiver-operating characteristic (ROC) curve \cite{hanley1982meaning}. In addition to these results, we can exploit the full joint distribution of our model to answer questions like \textit{what is the probability of jointly observing both edges $i \rightarrow j$ and $j \rightarrow i$?} This is equivalent to assign a value to the pair $(A_{ij}, A_{ji})$ from the set $\{(0,0),(0,1),(1,0),(1,1)\}$, that translates the edge prediction task into a classification problem. However, this problem becomes trivial if the model predicts all entries equal to $(0, 0)$: in this case we will get high performance just because of the high sparsity of the data. For this reason, we compute the accuracy only for entries in the test set that have at least one edge. For those, we predict the label associated to the highest probability among $[p_{01}, p_{10}, p_{11}]$, where these are computed by using \Crefrange{eqn:p01}{eqn:p11} with the inferred parameters. We then compute the accuracy between true and predicted labels, where a value equal to 1 means perfect recovery. As baselines, we use a uniform  random probability over the number of possible labels in the training set (RP), and the accuracy obtained by using as prediction the label with the highest relative frequency in the training set (MRF). The results are shown in \Cref{fig:auc_joints}, where we can observe a V-shape. Reciprocity equal to zero  (\rec\text{} $=0$) means the networks have no reciprocated edges, and higher its value higher the frequency of the label $(1,1)$. Thus, in the regime $0 \leq \mathsf{r}\leq 0.5$ the performance decreases because the problem becomes more difficult by reaching the point where labels have similar relative frequencies (MRF $\approx$ RP when \rec\text{} $=0.5$). In this scenario, \jcrep\text{} outperforms the baselines with a bigger gap as the reciprocity increases. When \rec\text{} $>0.5$ the problem becomes easier due to the increasing proportion of the label $(1,1)$. Here, predicting all entries equal to $(1, 1)$ results in higher performance (MRF). However, this is another trivial situation that should be ignored when analyzing the performance in edge prediction tasks.

\begin{figure}[htbp]
    \centering
   \makebox{ \includegraphics[width=0.7\linewidth]{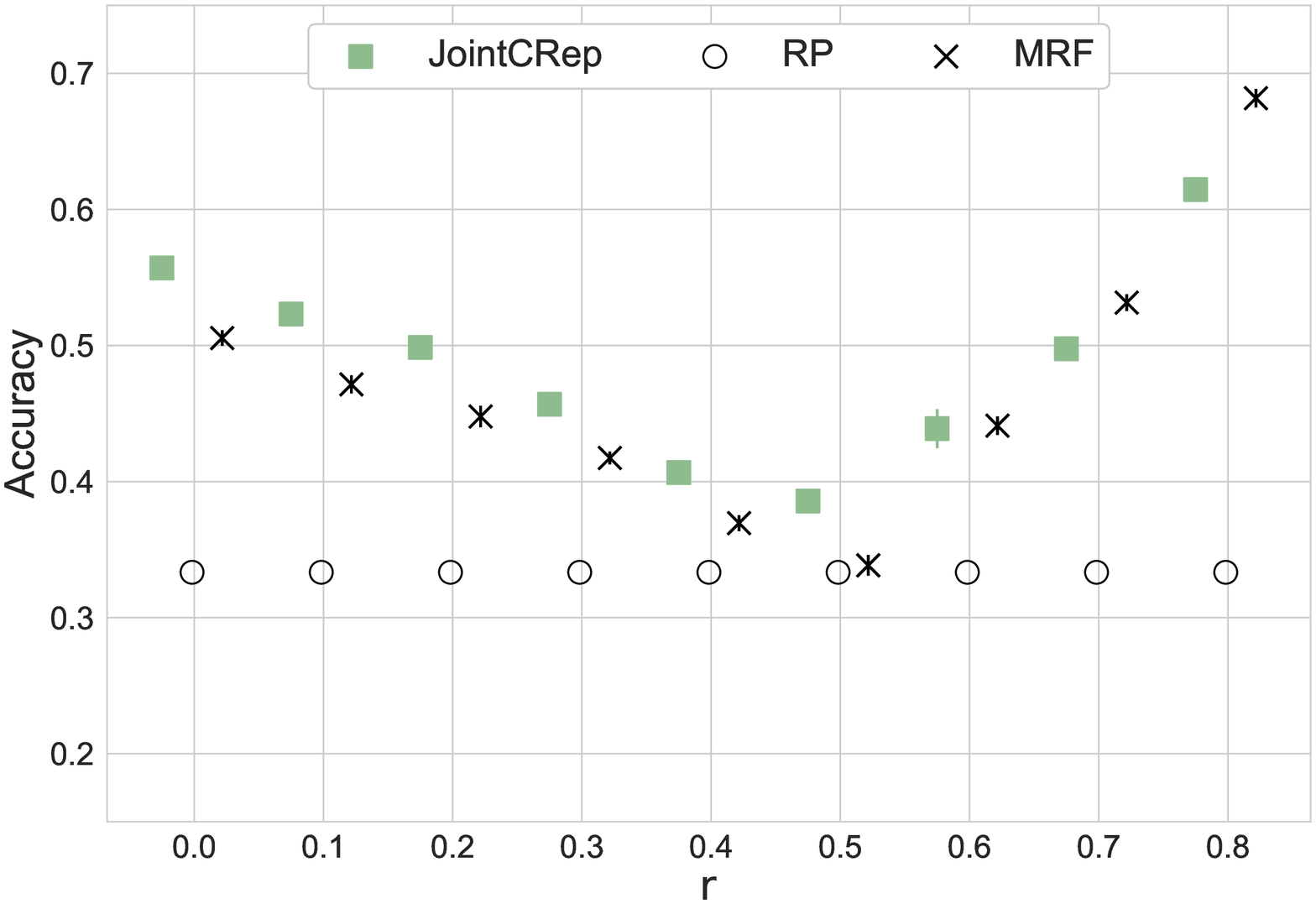}}
        \caption{ \label{fig:auc_joints} \textbf{Edge prediction with joint distributions in synthetic networks.} Synthetic networks with $N=1000$ nodes, $K=2$ overlapping communities, $\langle k \rangle = 20$ average degree, and different values of reciprocity \rec. Results are averages and standard deviations over $10$ synthetic networks and over $5$-folds of cross-validation test sets. Edge prediction performance is measured with accuracy, and as baselines we consider the uniform random probability (RP) and the maximum relative frequency (MRF).}
\end{figure}

\subsection{Reproducing network topological properties}\label{app:propertysample}
\Cref{fig:sampled_avg_degree} shows the performance of each model in reproducing the average degree in sampled networks. While \jcrep\text{} and \mt\text{} recover this feature despite the different values of reciprocity, \crep\text{} produces samples with a lower average degree than the one given in input as \rec\text{} increases. This happens because, in high reciprocity settings, \crep\text{} produces sampled networks with fewer edges but higher weights. Hence, while the average degree decreases, the weighted average degree better reflects the input feature (not shown here). 

\begin{figure}[htpb]
    \centering
    \makebox{\includegraphics[width=0.7\linewidth]{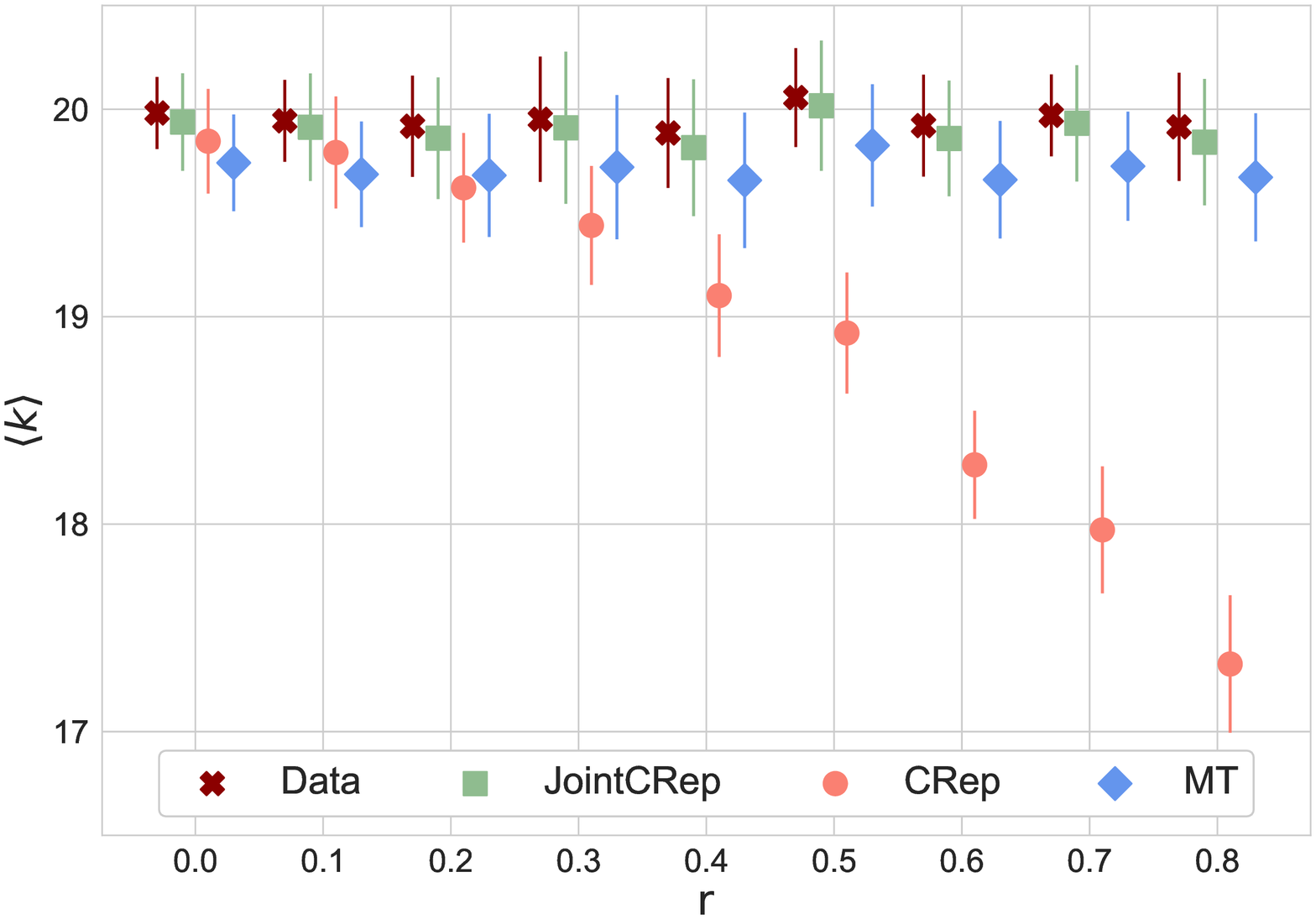}}
    \caption{\label{fig:sampled_avg_degree} \textbf{Average degree in sampled synthetic networks.}  Synthetic networks with $N=1000$ nodes, $K=2$ overlapping communities, $\langle k \rangle = 20$ average degree, and different values of reciprocity \rec. Results are empirical averages and standard deviations over $50$ samples of $10$ independent synthetic networks (five samples per input network). We measure the average degree and the dark red markers indicate the average on $10$ input networks.}
\end{figure}

\subsection{Analysis on synthetic data with different average degrees}\label{app:synAvgDegree}
In addition to the results provided in the manuscript, we also analyse synthetic networks with different values of average degree. \Cref{fig:res_AvgDegree} shows the performance of the models in community detection and edge prediction tasks, as well as in reproducing topological properties in sampled networks. Similar to the results in the manuscript, \jcrep\text{} follows the behaviour of \mt\text{} both in terms of CS and marginal AUC, for which performance improves as the average degree increases, as expected for community detection-only methods. On the other hand, \crep\text{} is only partially affected by the different values of average degree, as also shown in \cite{safdari2020generative}.  The plots highlight that the strength of \crep\text{} is not retrieving communities, rather its ability to predict missing edges by using the conditional distribution, regardless the average degree. In \Cref{fig:res_AvgDegree} we can also notice that even though \jcrep\text{} is affected by the average degree, its conditional AUC improves the marginal AUCs already when $\langle k \rangle = 4$, and it reaches good values from $\langle k \rangle = 10$. Moreover, \jcrep\text{} outperforms the other methods in recovering reciprocity in sampled networks across different values of average degree. Overall, these results suggest that \jcrep\text{} is a valuable tool also in networks with low-medium average degree providing good communities, reasonable edge predictions, and sampled networks with topological features that resemble that of the real data.

\begin{figure}[tb]
  \centering
  \begin{minipage}[t]{0.01\textwidth}
    \text{A}
  \end{minipage}
  \begin{minipage}[t]{0.48\textwidth}
    \includegraphics[width=\linewidth]{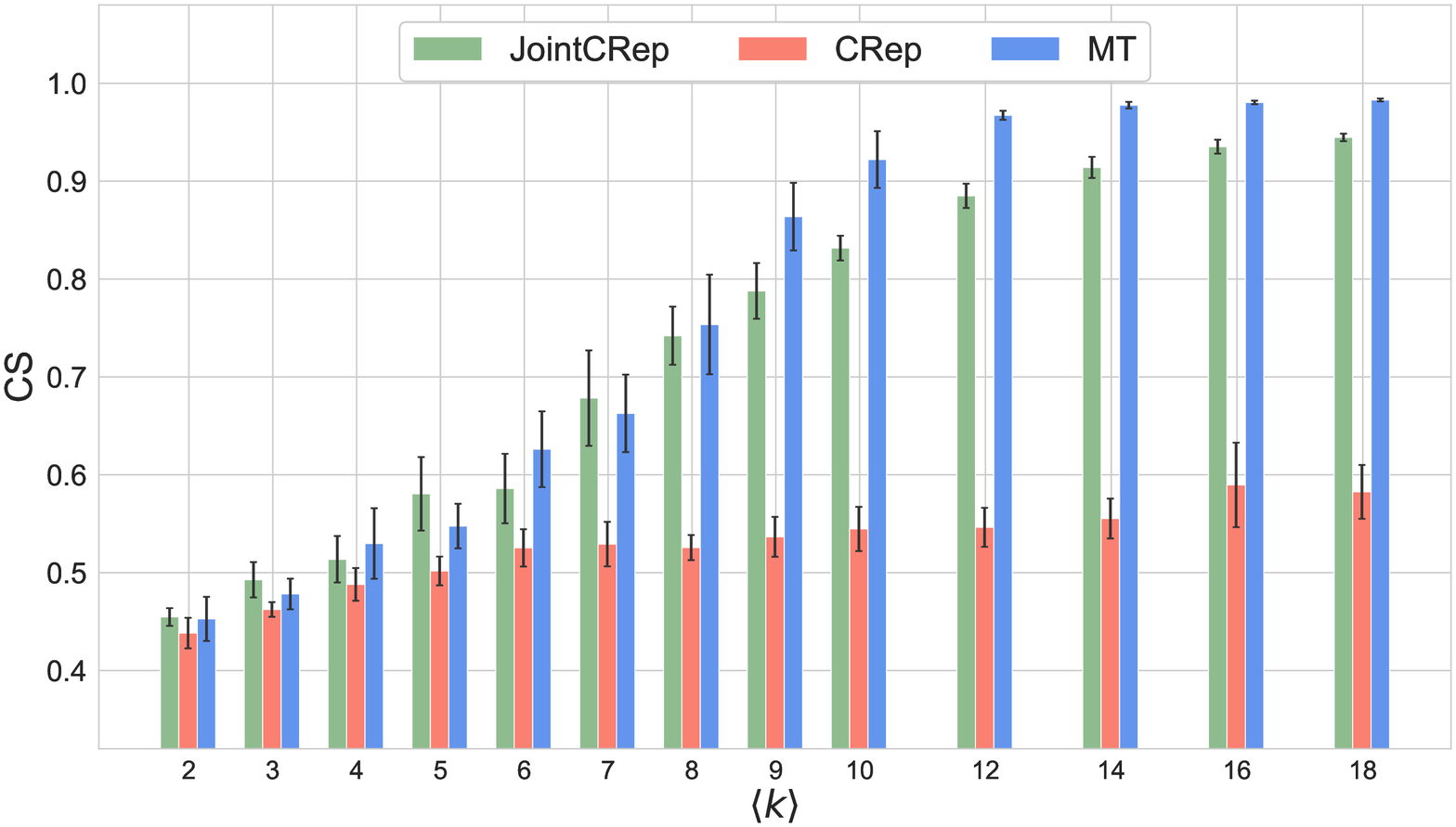}
  \end{minipage}\hfill
  \begin{minipage}[t]{0.01\textwidth}
    \text{B}
  \end{minipage}
  \begin{minipage}[t]{0.48\textwidth}
    \includegraphics[width=\linewidth]{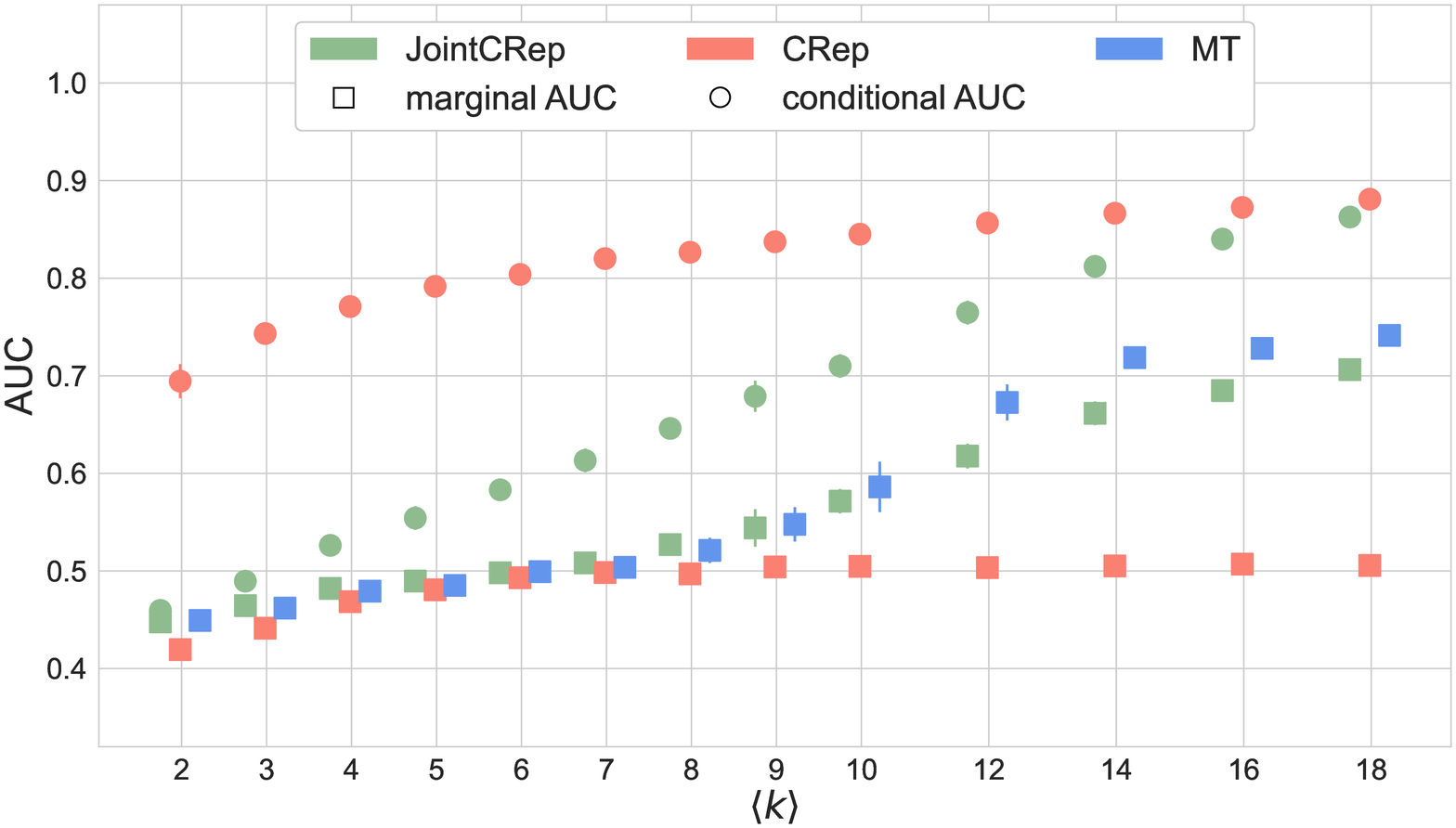}
  \end{minipage}
    \begin{minipage}[t]{0.01\textwidth}
    \text{C}
  \end{minipage}
  \begin{minipage}[t]{0.48\textwidth}
    \includegraphics[width=\linewidth]{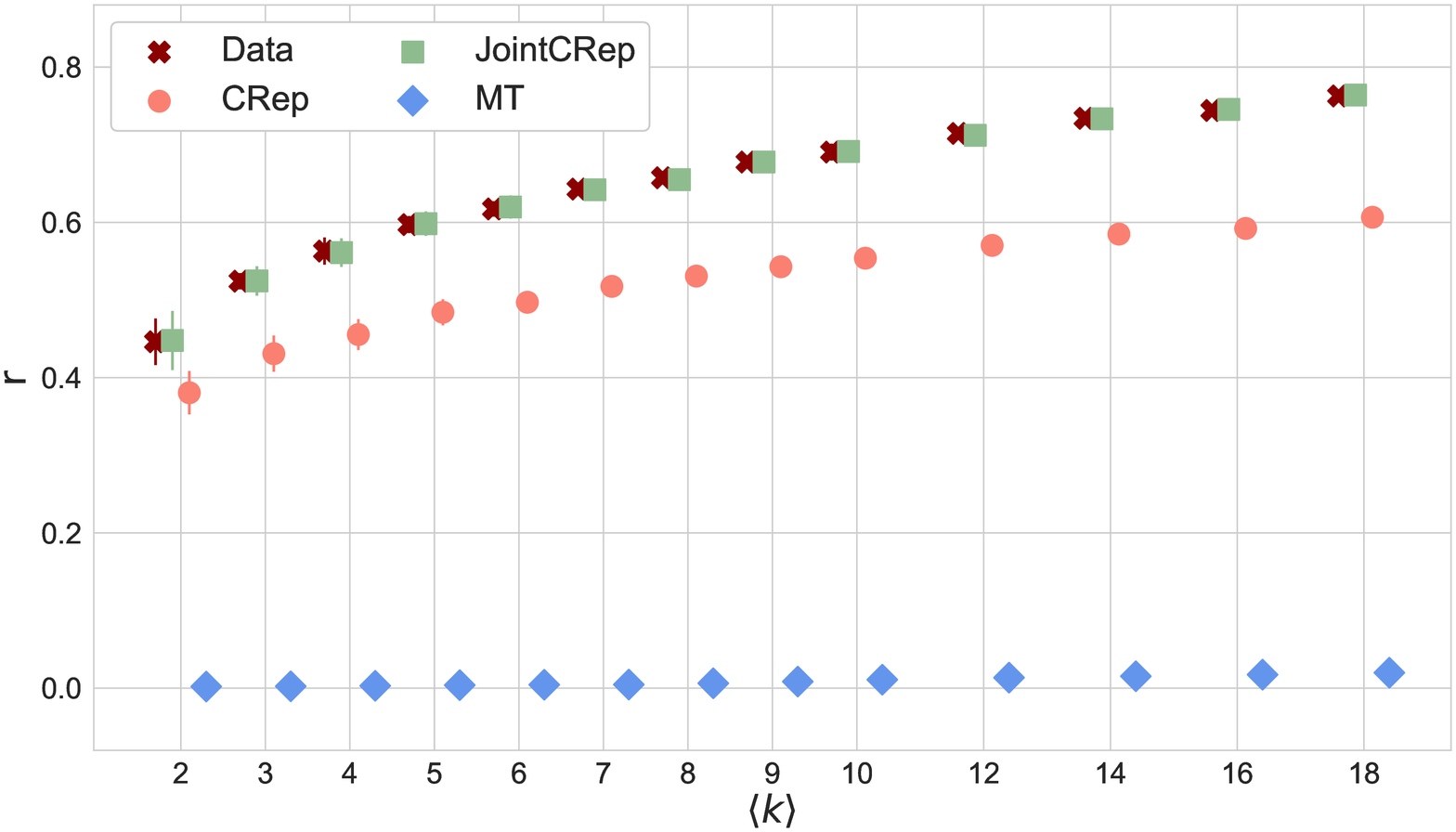}
  \end{minipage}\hfill
  \begin{minipage}[t]{0.01\textwidth}
    \text{D}
  \end{minipage}
  \begin{minipage}[t]{0.48\textwidth}
    \includegraphics[width=\linewidth]{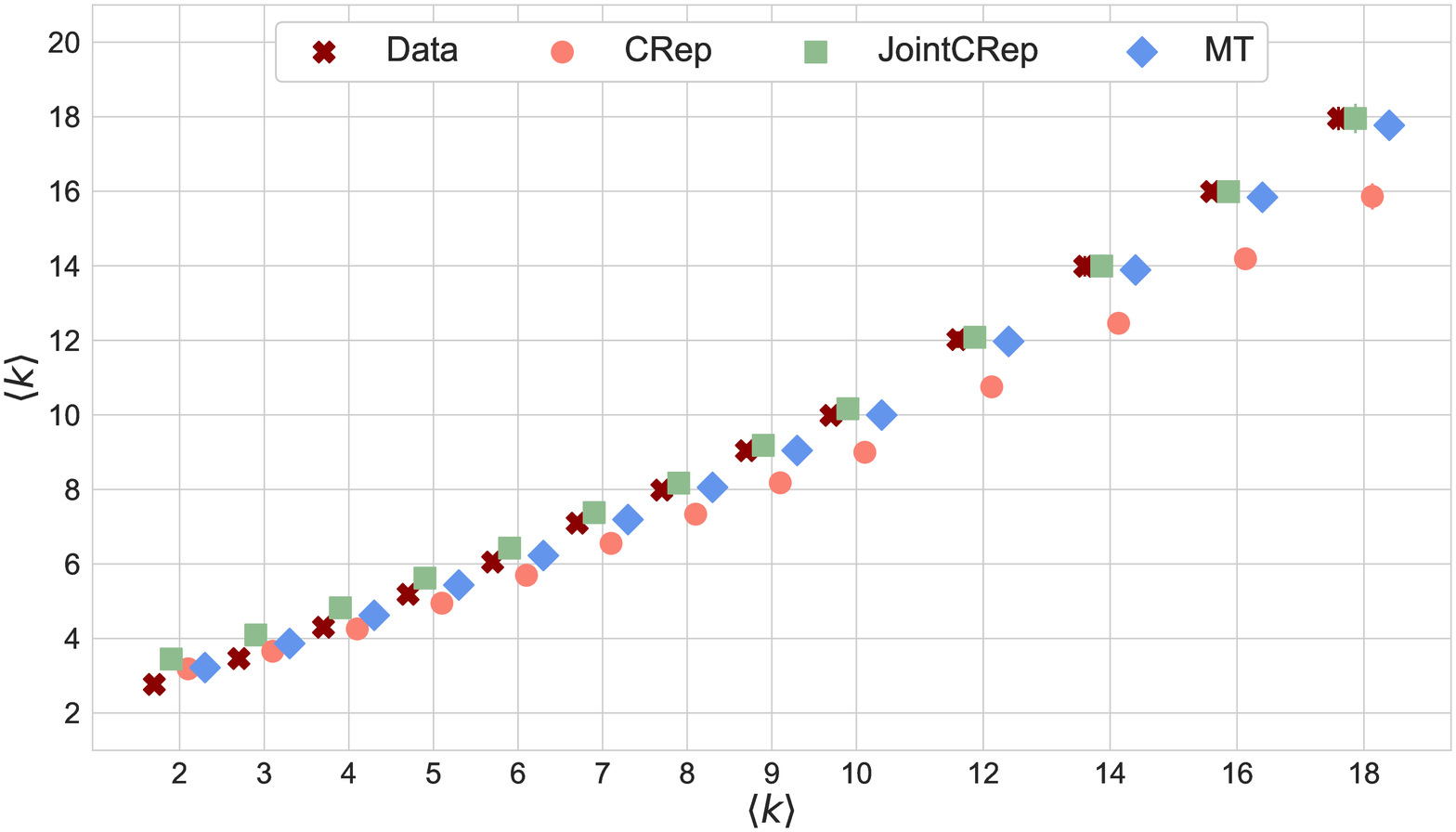}
  \end{minipage}
  \caption{\label{fig:res_AvgDegree} \textbf{Results on synthetic networks with different average degrees.}  Synthetic networks with $N=1000$ nodes, $K=2$ overlapping communities, pair-interaction parameter $\eta = 1000$, and different values of average degree $\langle k \rangle$. (A-B) Results are averages and standard deviations over 10 synthetic networks of (A) cosine similarity and (B) AUC. The latter measures the edge prediction performance over 5-folds of cross-validation test sets, and the baseline is the random value 0.5. (C-D) Results are empirical averages and standard deviations over 50 samples of 10 independent synthetic networks (five samples per input network). We measure (C) the reciprocity and (D) the average degree, and the dark red markers indicate the average on 10 input networks.}
 \end{figure}

\section{Results on real-world datasets}
\subsection{Loss functions} \label{app:lossfunctions}
In addition to the AUC, we use the Log Loss (or Binary Cross-Entropy) and the L1 Loss (or Mean Absolute Error) to measure the performance of the methods in edge prediction and network reconstruction. The Log Loss for binary classification is defined as
\begin{equation}
- \f{1}{M}\sum_{i,j} \rup{A_{ij} \, \log\,P(A_{ij}) \, + \, (1-A_{ij}) \, \log\,\bup{1-P(A_{ij})}}\quad,
\end{equation}
where $A_{ij}$ indicates the entry of the adjacency matrix, $P(A_{ij})$ denotes the probability of the existence of the edge, and $M$ is the total number of edges.\\
 Instead, the L1 Loss is given by 
\begin{equation}
\f{1}{M} \sum_{i,j} | A_{ij}  - P(A_{ij})| \quad .
\end{equation}
For both metrics, lower values indicate better performance and a loss of 0 denotes perfect predictions. While the Log Loss does not have an upper bound, the L1 Loss is equal to 1 in the worst-case scenario of predicting every existing edge with probability $P(A_{ij}=1)=0$ and every non-existing edge with probability $P(A_{ij}=1)=1$. Moreover, they differ in the extent to which they penalize mistakes: the Log Loss is more sensitive to large disagreements between true and predicted values than the L1 Loss. This means that the Log Loss prefers predictions with more mistakes of low magnitude than predictions with fewer mistakes but of larger magnitude. 

\subsection{Analysis of a high-school social network}\label{app:HST11_app}
\Cref{tab:CV_highschool} displays the results for the edge prediction task in the high-school social network. \crep\text{} performs the best both in terms of AUC and Log Loss when using the conditional probabilities. On the other hand, \jcrep\text{} is the best when considering the L1 Loss. This is explained by the behaviour of our model that tends to predict fewer edges with more intensity, differently to the other models which predict more edges with low-medium probabilities. As a remark, the dataset presents an average degree $\langle k \rangle=6.45$ and it is highly sparse. This feature makes this task hard because there is only little information in input when considering 5-fold cross-validation splits, and some folds may result in unreliable results. Nevertheless, results show that all models are performing reasonably well at this task given this sparse regime. \\
Comparing the communities inferred by the various methods,  \Cref{tab:Qnicosia} shows the overlapping modularity obtained for the partitions of \Cref{fig:HST11_community_soft_u} for various aggregation functions.  \Cref{tab:metricsReconstruction_HS} reports the penalties for the network reconstruction task, showing that  \jcrep\text{} has best performance in terms of both Log Loss and L1 Loss. Finally, \Cref{tab:high_school_samples} shows the topological properties in the high-school social network and its sampled networks, showing how  \jcrep\text{}  achieves on average values that are more similar to those observed on the input data.

\begin{table*}[htpb]
\caption{\label{tab:CV_highschool} \textbf{Edge prediction in the high-school social network.} Results are averages and standard deviations over 5-folds of cross-validation test sets. Edge performance is measured  with three different metrics $\mathcal{F}$: AUC, Log Loss, and L1 Loss. The AUC measures the probability that a randomly selected edge has higher expected value than a randomly selected non-existing edge, and the baseline is the random value 0.5. The Log Loss and the L1 Loss are penalty measures defined in \Cref{app:lossfunctions}, that quantify the difference between two input networks by taking into account the probability of the existence of an edge and computing a penalty for each mistake in predicting the observed value. The metrics are computed by using either the marginal probability $P(A_{ij} | \Theta)$ or the conditional probability $P(A_{ij} | A_{ji}, \Theta)$. Note that the last is not available for \mt\text{} because the conditional and marginal distributions coincide. The best performance for each metric is in bold.}
\centering
\begin{adjustbox}{angle=0}
\resizebox{1\textwidth}{!}{%
{\renewcommand{\arraystretch}{1.}
\setlength\arrayrulewidth{0.7pt}
\begin{tabular}{*{6}{c}}            
\toprule
  & \multicolumn{3}{c}{$P(A_{ij} | \Theta)$}     
    & \multicolumn{2}{c}{$P(A_{ij} | A_{ji}, \Theta)$}    \\
                       \cmidrule(lr){2-4}                  
    \cmidrule(lr){5-6}
                 $\mathcal{F}$                  & \jcrep                                               & \crep                         & \mt       & \jcrep               & \crep       \\ \midrule  
AUC                    & 0.610  $\pm$ 0.061                        & 0.650  $\pm$ 0.109               & \textbf{0.668 $\pm$ 0.111}             &   0.626 $\pm$ 0.073             & \textbf{0.786 $\pm$ 0.055}      \\
Log Loss              & 0.825  $\pm$  0.191                 & \textbf{0.678     $\pm$  0.190}           & 0.726 $\pm$   0.284        &   0.820 $\pm$    0.213           &  \textbf{0.492 $\pm$ 0.124}   \\
L1 Loss              & 0.133    $\pm$ 0.014			 & 0.139 $\pm$	 0.015	      & \textbf{0.132 $\pm$    0.026}       &   \textbf{0.122 $\pm$ 0.019}               &  0.125   $\pm$ 0.014    \\
\bottomrule                  
\end{tabular}}}
\end{adjustbox}
\end{table*}

\begin{table*}[htpb]
\caption{\label{tab:Qnicosia} \textbf{Modularity for the high-school social network.} Values are computed using the overlapping formulation as in \citet{nicosia2009extending}, and $\mathcal{F}$ denotes the aggregation function considered in each row. We use the mixed-membership partitions determined by the matrix $u$ inferred by \jcrep, \crep, and \mt. Results are similar for the matrix $v$.}
\centering
\begin{adjustbox}{angle=0}
\resizebox{0.6\textwidth}{!}{%
{\renewcommand{\arraystretch}{1.2}
\setlength\arrayrulewidth{0.7pt}
\begin{tabular}{*{4}{c}}
\toprule
                 $\mathcal{F}$                  & \jcrep                                               & \crep                           & \mt                     \\ \midrule
Mean                    & 0.74                                             & 0.72                    & 0.75                                                   \\
Max                   & 0.65    			 & 0.62 	 	      & 0.48      \\
Product           &  0.55 & 0.53 &0.73 \\
\bottomrule                                    
\end{tabular}}}
\end{adjustbox}
\end{table*}

\begin{table*}[htpb]
\caption{\label{tab:metricsReconstruction_HS} \textbf{High-school network reconstruction: comparison between true and reconstructed networks.} $\mathcal{F}$ denotes the function considered in each row, defined in \Cref{app:lossfunctions}. The metrics are computed by using either the marginal probability $P(A_{ij} | \Theta)$ or the conditional probability $P(A_{ij} | A_{ji}, \Theta)$ of each method with the inferred parameters. Note that the last is not available for \mt\text{} because the conditional and marginal distributions coincide. The best performance for each metric is in bold.}
\centering
\begin{adjustbox}{angle=0}
\resizebox{0.9\textwidth}{!}{%
{\renewcommand{\arraystretch}{1.2}
\setlength\arrayrulewidth{0.7pt}
\begin{tabular}{*{6}{c}}
\toprule
  & \multicolumn{3}{c}{$P(A_{ij} | \Theta)$}     
    & \multicolumn{2}{c}{$P(A_{ij} | A_{ji}, \Theta)$}    \\
                       \cmidrule(lr){2-4}                  
    \cmidrule(lr){5-6}
            $\mathcal{F}$                  & \jcrep                                            & \crep                         & \mt      & \jcrep              & \crep       \\ \midrule
Log Loss                    & \textbf{0.144}                                           & 0.307                    & 0.165             &      \textbf{0.128}   &  0.185                                                   \\
L1 Loss                   & \textbf{0.093}    			 & 0.137 	 	      & 0.106      &   \textbf{0.077}         & 0.120      \\
\bottomrule                                    
\end{tabular}}}
\end{adjustbox}
\end{table*}

\begin{table*}[htpb]
\caption{\label{tab:high_school_samples} \textbf{Topological properties in the high-school social network and its sampled networks.} Results are averages and standard deviations over five samples. We measure the number of nodes $N$, the number of edges $M$, the average degree $\langle k \rangle$, the reciprocity \rec, and the clustering coefficient $cc$.}
\centering
\setlength{\tabcolsep}{10pt}
\begin{adjustbox}{angle=0}
\resizebox{0.9\linewidth}{!}{%
{\renewcommand{\arraystretch}{1.2}
\setlength\arrayrulewidth{0.7pt}
\begin{tabular}{*{6}{c}}
\toprule
                      & $N$    & $M$   &  $\langle k \rangle$  & \rec & $cc$                                        \\ \midrule
 Data   & 31                     & 100                     & 6.45                  & 0.52                  & 0.38 \\
 \jcrep & $ 30.8 \pm 0.45  $ & $ 90.8 \pm 6.76$   & $ 5.89 \pm 0.38 $ & $0.47 \pm 0.06 $ & $0.20 \pm 0.02$ \\
 \crep  & $ 30.6 \pm 0.55  $ & $ 77.8   \pm 12.79  $ & $5.08   \pm 0.81 $ & $0.49 \pm 0.08$  & $0.11 \pm 0.05 $ \\
 \mt     & $31 \pm 0  $ & $ 79.8      \pm 2.05  $ & $5.15   \pm 0.13 $ & $ 0.21 \pm 0.04 $ & $0.24 \pm 0.04$ \\           
\bottomrule                                    
\end{tabular}}}
\end{adjustbox}
\end{table*}

\subsection{Analysis of a vampire bat network}
\Cref{tab:CV_vampire} shows  results for edge prediction tasks using a 5-fold cross-validation routine. Similarly to the high-school dataset, all the models obtain good performance given the sparse regime, although values were slightly better in the high-school case. Also in this case,  \jcrep\text{} achieves best results in terms of L1 Loss, and \crep\text{} is more robust in terms of Log Loss. \Cref{tab:metricsReconstruction_vampire} reports the quantitative results for the network reconstruction of the vampire bat dataset, in terms of Log and L1 Loss. Here, \jcrep\text{} outperforms the other methods as also shown in \Cref{fig:vampire_bat_adjacency}. 

\begin{table*}[htpb]
\caption{\label{tab:CV_vampire} \textbf{Edge prediction in the vampire bat network.} Results are averages and standard deviations over 5-folds of cross-validation test sets. Edge performance is measured  with three different metrics $\mathcal{F}$: AUC, Log Loss, and L1 Loss. The AUC measures the probability that a randomly selected edge has higher expected value than a randomly selected non-existing edge, and the baseline is the random value 0.5. The Log Loss and the L1 Loss are penalty measures defined in \Cref{app:lossfunctions}, that quantify the difference between two input networks by taking into account the probability of the existence of an edge and computing a penalty for each mistake in predicting the observed value. The metrics are computed by using either the marginal probability $P(A_{ij} | \Theta)$ or the conditional probability $P(A_{ij} | A_{ji}, \Theta)$. Note that the last is not available for \mt\text{} because the conditional and marginal distributions coincide. The best performance for each metric is in bold.}
\centering
\begin{adjustbox}{angle=0}
\resizebox{1\textwidth}{!}{%
{\renewcommand{\arraystretch}{1.}
\setlength\arrayrulewidth{0.7pt}
\begin{tabular}{*{6}{c}}
\toprule
  & \multicolumn{3}{c}{$P(A_{ij} | \Theta)$}     
    & \multicolumn{2}{c}{$P(A_{ij} | A_{ji}, \Theta)$}    \\
                       \cmidrule(lr){2-4}                  
    \cmidrule(lr){5-6}
                 $\mathcal{F}$                  & \jcrep                                               & \crep                         & \mt       & \jcrep               & \crep       \\ \midrule  
AUC                    & \textbf{0.687  $\pm$ 0.078}                        & 0.627  $\pm$ 0.079              & 0.629 $\pm$ 0.073            & 0.715 $\pm$ 0.098              & \textbf{0.772 $\pm$ 0.063}      \\
Log Loss              & 1.514  $\pm$  0.282                 &\textbf{0.961     $\pm$  0.196}           & 1.804 $\pm$   0.147        & 1.391 $\pm$   0.269           &  \textbf{0.812 $\pm$ 0.261}   \\
L1 Loss              &\textbf{0.277    $\pm$ 0.031}		 & 0.340 $\pm$	 0.014	      & 0.291  $\pm$    0.020        &   \textbf{0.229 $\pm$ 0.021}               &  0.296   $\pm$ 0.008    \\
\bottomrule                                    
\end{tabular}}}
\end{adjustbox}
\end{table*}

\begin{table*}[htbp]
\caption{\label{tab:metricsReconstruction_vampire} \textbf{Vampire bat network reconstruction: comparison between true and reconstructed networks.} $\mathcal{F}$ denotes the function considered in each row, defined in \Cref{app:lossfunctions}. The metrics are computed by using either the marginal probability $P(A_{ij} | \Theta)$ or the conditional probability $P(A_{ij} | A_{ji}, \Theta)$ of each method with the inferred parameters. Note that the last is not available for \mt\text{} because the conditional and marginal distributions coincide. The best performance for each metric is in bold.}
\centering
\begin{adjustbox}{angle=0}
\resizebox{0.9\textwidth}{!}{%
{\renewcommand{\arraystretch}{1.}
\setlength\arrayrulewidth{0.7pt}
\begin{tabular}{*{6}{c}}
\toprule
  & \multicolumn{3}{c}{$P(A_{ij} | \Theta)$}     
    & \multicolumn{2}{c}{$P(A_{ij} | A_{ji}, \Theta)$}    \\
                       \cmidrule(lr){2-4}                  
    \cmidrule(lr){5-6}
                 $\mathcal{F}$                  & \jcrep                                               & \crep                         & \mt       & \jcrep               & \crep       \\ \midrule   
Log Loss                    &  \textbf{0.173}                                             & 0.466                    & 0.302             &      \textbf{0.159}   &  0.414                                                   \\
L1 Loss                   &  \textbf{0.110}    			 & 0.308 	 	      & 0.207      &   \textbf{0.103}         & 0.275      \\
\bottomrule                                    
\end{tabular}}}
\end{adjustbox}
\end{table*}

\end{widetext}
\end{document}